\documentclass[twocolumn,english,aps,jcp,superscriptaddress,showpacs,floatfix]{revtex4}
\usepackage{graphicx}
\usepackage{amsmath,amssymb,amsfonts}
\usepackage{natbib}

\usepackage{babel}
\begin{document}

\title{Efficient and accurate surface hopping for long time nonadiabatic quantum dynamics.}

\author{Aaron Kelly}
\author{Thomas E. Markland}
\email{tmarkland@stanford.edu}
\affiliation{Department of Chemistry, Stanford University, Stanford, CA 94305, USA}

\date{\today{}}

\begin{abstract}
The quantum-classical Liouville equation offers a rigorous approach to nonadiabatic quantum dynamics based on surface hopping type trajectories. However, in practice the applicability of this approach has been limited to short times owing to unfavorable numerical scaling. In this paper we show that this problem can be alleviated by combining it with a formally exact generalized quantum master equation treatment. This allows dramatic improvements in the efficiency of the approach in nonadiabatic regimes, making it computationally tractable to treat the quantum dynamics of complex systems for long times. We demonstrate our approach by applying it to a model of condensed phase charge transfer where our method is shown to be numerically exact in regimes where fewest-switches surface hopping and mean field approaches fail to obtain the either the correct rates or long-time populations.
\end{abstract}

\maketitle

\section{Introduction} 
\label{sec:intro}

One of the central problems in condensed phase dynamics concerns the development of accurate and efficient methods for the treatment of charge and energy transport. These processes are inherently quantum mechanical and underpin many of the forefront problems in chemistry, biology and materials science, such as catalysis involving electron transfer \cite{Marcus64,Marcus85}, proton transfer \cite{bell73,hanna05}, or proton-coupled electron transfer (PCET) \cite{cukier98,mayer04,shs08} and electronic excitation energy transfer \cite{engel07,lee07,collini09,ishizaki09}. Due to the poor scaling of exact quantum dynamics with both time and system dimensionality these approaches are only feasible for a small class of models or for specific forms of the potential \cite{tanimura89, tanimura91, makri92, makarov94, makri94a, makri94b, egger94,shi03a,muhlbacher08, golosov99,cohen11}. Hence, in complex systems, a multi-tiered approach is typically employed in which the most important or most quantum mechanical degrees of freedom are treated as a quantum subsystem while the remaining degrees of freedom are designated as the bath which can be treated at a lower level of theory, such as by classical mechanics \cite{kapral06,Tully12,berkelbach12}. Adopting such a scheme has led to the development of a series of trajectory-based methods that allow the treatment of quantum dynamics in large condensed phase systems for general potentials. The most computationally efficient of these are mean-field type theories, such as Ehrenfest theory or those arising from linearization of the quantum propagator \cite{mclachlan64,sun98,poulsen03,shi03c,bonella05,kim-map08,bonella10}. Although these methods are computationally efficient they fail to provide an accurate description of the approach to equilibrium in relaxation processes when the subsystem levels are non-degenerate, and they cannot account for quantum coherence in the bath degrees of freedom\cite{Miller12}. These inaccuracies are emphasized when the subsystem and the bath are strongly coupled. Higher accuracy can be achieved by foregoing linearization of the quantum subsystem degrees of freedom, thus obtaining a partially linearized approach \cite{miller70,dunkel08,huo12,kapral99}. However, simulations based on these techniques are limited to short time-scales due to undesirable scaling of their computation cost with evolution time.

Another important consideration in trajectory-based approaches is the basis in which the quantum subsystem is represented. Typically in model problems the Hamiltonian is defined in the diabatic basis. However, electronic structure calculations produce adiabatic surfaces and hence combination with on-the-fly electronic structure is dramatically aided by working in the adiabatic representation. The trajectory-based fewest switches surface hopping (FSSH) \cite{tully71,tully90}, and its variants \cite{bittner95,prezhdo97,subotnik11a,shuskov12}, have thus become widely used. However, FSSH lacks a rigorous derivation, suffers from decoherence problems, and has also recently been shown to be deficient in treating isolated electron transfer due to its inability to obtain the correct scaling of Marcus' golden rule rate \cite{subotnik11b}. Several schemes have been formulated in order to attempt to solve these issues however, due to the lack of a formal derivation they simply provide {\it ad hoc} corrections \cite{bedard-hearn05,subotnik12}. 

A rigorous method to generate dynamics in the adiabatic representation is provided by the momentum-jump (MJ) solution to the quantum-classical Liouville equation (QCLE) \cite{mackernan02,mackernan08}. The MJ solution yields a surface-hopping type algorithm in which evolution on adiabatic surfaces is interspersed with instantaneous jumps between surfaces. For weakly nonadiabatic systems, where the average number of jumps between surfaces is small, the MJ solution provides a highly accurate and computationally tractable approach \cite{kelly10}. However, in strongly nonadiabatic regimes, where multiple jumps between surfaces are likely to occur, the weights accumulated upon making jumps give rise to a very rapidly growing statistical error at long times. Hence, convergence of the population dynamics by the direct application of these methods is intractable at the times required for the population dynamics to fully decay. 

It would therefore be highly advantageous if the rich information embodied in the short time trajectories generated from highly accurate partially linearized methods could be used to generate long time population dynamics. The generalized quantum master equation (GQME) formalism offers a method to achieve this by allowing the problem of obtaining dynamics at arbitrarily long times to be exactly mapped onto the calculation of the memory kernel \cite{nakajima58,zwanzig60}. In cases where the range of frequencies present in the bath is larger than that of the subsystem, as is typically the case in condensed phase systems, the memory kernel can decay many orders of magnitude more quickly than the population dynamics. This observation has previously been used to improve the efficiency of real-time path integral approaches \cite{cohen11} and second order cumulant expansion techniques \cite{meier99,xu02,kleinekathofer04}. It has also been leveraged in order to improve the accuracy of fully linearized techniques that can be applied to a much wider set of Hamiltonians \cite{shi04a,shi04b}. Here, we show how the GQME formalism can be exploited to generate long-time dynamics from a partially linearized approach that, due to poor scaling with time, cannot be converged at long times when used directly. 

In particular, we combine the MJ solution to the QCLE in the adiabatic basis with the generalized quantum master equation formalism to yield a surface hopping approach that is both efficient and accurate. By making such a combination we invoke no additional approximations to the MJ solution, while significantly reducing the computational cost compared with direct MJ dynamics. The resulting MJ-GQME method can therefore be applied to provide accurate long time dynamics in nonadiabatic regimes where FSSH and linearized approaches are inaccurate and direct MJ dynamics cannot be converged. Since it is based on semi-classical trajectories the MJ-GQME approach is flexible enough to be applied to general Hamiltonians which are not amenable to other accurate methods which typically require specific forms of the Hamiltonian such as harmonic environments or bi-linear system-bath coupling. We demonstrate our approach by applying it to the spin boson problem in regimes where typical trajectory-based approaches to quantum dynamics fail.

\section{Theory} 
\label{sec:theory}

We first describe an adiabatic basis MJ solution to the QCLE, which yields an ensemble of surface-hopping type trajectories. We then show how this solution can be combined with the GQME formalism in order to yield a method which allows one to perform long-time nonadiabatic dynamics simulations. 

\subsection{Surface-hopping in the adiabatic basis}
\label{subsec:MJ}

Consider a system in which only a small subset of the degrees of freedom behave quantum mechanically to a significant degree, for example electronic degrees of freedom, or light atoms such as protons. We denote this set of degrees of freedom as the quantum subsystem (or simply as the subsystem). The remainder of the system, typically the heavy atoms, is referred to as the bath and is assumed to behave essentially classically. This allows one to write the total Hamiltonian in the following form, 
\begin{equation} \label{eq:H_tot}
\hat{H} = \hat{H}_s + \hat{H}_b + \hat{H}_{sb},\end{equation} 
where the subscripts $s, b,$ and $sb$ refer to the subsystem, the bath, and the system-bath coupling,  respectively.  The quantum-classical Liouville equation \cite{kapral99},
\begin{eqnarray} \label{eq:qcle}
  && \frac{\partial }{\partial t}\hat{\rho}_W (X, t) =-i  {\mathcal L}\hat{\rho}_W (X,t),
\end{eqnarray}
describes the time evolution of the density matrix $\hat{\rho}_W(X,t)$, which is a quantum mechanical operator that depends on the classical phase space variables $X=(R,P)=(R_1,R_2,...,R_{N_b},P_1,P_2,...,P_{N_b})$ of the bath. The quantum-classical Liouville (QCL) operator is
\begin{equation}\label{eq:qcl_op}
i{\mathcal L} \cdot = \frac{i}{\hbar}[\hat{H}_W,\cdot] - \frac{1}{2}(\{\hat{H}_W,\cdot\}
-\{\cdot,\hat{H}_W\}),
\end{equation}
where $[\cdot,\cdot]$ is the commutator, and $\{\cdot,\cdot\}$ is the Poisson bracket in the phase space of the environmental variables. The subscript $W$ refers to the partial Wigner transform over the environmental degrees of freedom in the system. The partial Wigner transform for an arbitrary operator, $\hat{A}$, is
\begin{eqnarray} \label{eq:wigner}
\hat{A}_{W}(R,P) = \int dZ e^{-i P \cdot Z} \langle R + \frac{Z}{2} | \hat{A} | R -\frac{Z}{2}\rangle.  
\end{eqnarray}

In the adiabatic basis the QCLE leads to a surface-hopping-type solution where segments of classical evolution are interspersed with energy-conserving quantum transitions\cite{kapral99,mackernan02}. The QCL operator may be written in this basis as a sum of diagonal (adiabatic) and off-diagonal (nonadiabatic) parts,
\begin{equation}\label{eq:QCL_adi}
i\mathcal{L}_{\alpha \alpha' \beta \beta'} = i\mathcal{L}^d_{\alpha \alpha'}\delta_{\alpha \beta} \delta_{\alpha' \beta'} - \mathcal{J}_{\alpha \alpha' \beta \beta'}.
\end{equation} 
The adiabatic part of the operator is explicitly given by
\begin{equation} \label{eq:QCL_diag} i\mathcal{L}^d_{\alpha \alpha'}  = i( \omega_{\alpha \alpha'}(R) + L_{\alpha \alpha'}),\end{equation} where
\begin{equation} \label{eq:L_class}
iL_{\alpha \alpha'} = \frac{P}{M} \cdot \frac{\partial}{\partial R} + \frac{1}{2} (F_W^{\alpha}+F_W^{\alpha'}) \cdot \frac{\partial}{\partial P}
\end{equation}  
is the classical propagator on the $(\alpha \alpha')$ adiabatic surface, $F_W^{\alpha} = -\langle \alpha;R|\frac{\partial \hat{H}_W}{\partial R}|\alpha;R\rangle$ is the Hellmann-Feynman force on adiabatic state $\alpha$, and $\omega_{\alpha \alpha'}(R) = \frac{1}{\hbar}(E_{\alpha}(R)-E_{\alpha'}(R))$ is the dynamical phase.

The nonadiabatic part of the QCL operator, $\mathcal{J}$, produces quantum transitions between adiabatic surfaces and associated momentum transfer with the environmental degrees of freedom.
\begin{eqnarray}\label{eq:Jump_op}
\nonumber \mathcal{J}_{\alpha \alpha' \beta \beta'}& = &-\frac{P}{M} \cdot d_{\alpha \beta} \left( 1 + \mathcal{S}_{\alpha \beta}\frac{\partial}{\partial P}\right)\delta_{\alpha' \beta'} \\
&& -\frac{P}{M} \cdot d^{*}_{\alpha' \beta'} \left( 1 + \mathcal{S}^{*}_{\alpha' \beta'}\frac{\partial}{\partial P}\right)\delta_{\alpha \beta}
\end{eqnarray} Here $d_{\alpha \beta} = \langle \alpha;R|\frac{\partial }{\partial R}|\beta;R\rangle$ are the matrix elements of the nonadiabatic coupling vector, and $\mathcal{S}_{\alpha \beta} = \hbar \omega_{\alpha \beta} d_{\alpha \beta} (\frac{P}{M} \cdot d_{\alpha \beta})^{-1}$.

In order to arrive at a surface hopping-type representation the overall propagator is divided into a string of $N_t$ short time propagators of length $\delta$, such that $t=N_t \delta$. Denoting pairs of quantum states $(\alpha \alpha')$ by a single index, 
$s = \mathcal{N}\alpha + \alpha'$ where $\mathcal{N}$ is the dimensionality of the subsystem Hilbert space, the matrix elements of the total propagator may be written as
\begin{equation} \label{eq:total_prop}
\left(e^{i\mathcal{L}t}\right)_{s_0 s_N} = \sum_{s_1 \cdots s_{N-1}} \prod_{j=1}^{N} \left(e^{i\mathcal{L}\delta} \right)_{s_{j-1} s_j}.
\end{equation}
A Trotter factorization is then applied to each short time propagator in the above product, in order to separate the adiabatic and nonadiabatic parts,
\begin{equation}\label{eq:trotter_1}
\left(e^{i\mathcal{L}\delta}\right)_{s s'} =  \left(e^{i\mathcal{L}^d \delta / 2} \right)_{s}\left(e^{-\mathcal{J}\delta} \right)_{s s'}\left(e^{i\mathcal{L}^d \delta / 2} \right)_{s'} + \mathcal{O} (\delta^3).
\end{equation} A further Trotter factorization is then applied to each of the short-time nonadiabatic propagators, which yields \cite{mackernan08}
\begin{equation}\label{eq:trotter_2}
e^{-\mathcal{J}\delta} = Q_1(\delta) \left( 1 + C \frac{\partial}{\partial P} \right) + \mathcal{O}(\delta^2). 
\end{equation} The explicit forms of the matrices $Q_1$ and $C$ for a two level system are 
\begin{eqnarray}\label{eq:Q_1}\nonumber &&Q_1(\delta)= \\\nonumber &&\left(\begin{array}{cccc}\cos^2{a} & -\cos{a}\sin{a}  & -\cos{a}\sin{a}  & \sin^2{a} \\\cos{a}\sin{a} & \cos^2{a} & -\sin^2{a}  & -\cos{a}\sin{a} \\ \cos{a}\sin{a} &  -\sin^2{a} & \cos^2{a} & \cos{a}\sin{a} \\\sin^2{a} & \cos{a}\sin{a} & \cos{a}\sin{a} & \cos^2{a}\end{array}\right),\\
\end{eqnarray} 
where $a = \frac{P}{M} \cdot d_{21} \delta$, and
\begin{eqnarray}\label{eq:C}C=\left(\begin{array}{cccc} 0 & \mathcal{S}_{12} & \mathcal{S}_{12}  & 2\mathcal{S}_{12} \\ \mathcal{S}_{21} & 0 & 0 & \mathcal{S}_{12} \\\mathcal{S}_{21} & 0 & 0 & \mathcal{S}_{12} \\ 2\mathcal{S}_{21} & \mathcal{S}_{21}  & \mathcal{S}_{21}  & 0 \end{array}\right).\\\nonumber
\end{eqnarray} 

Other than the Trotter factorization of the short-time propagators, no further approximations have been made to the QCLE thus far. However due to the linear term $(1 + C \frac{\partial}{\partial P})$ in Eq.(\ref{eq:trotter_2}), the action of the short-time nonadiabatic propagators is not straightforward to evaluate. Hence, at this point one may invoke the momentum-jump (MJ) approximation \cite{kapral99,mackernan02,mackernan08}, where the linear term in Eq.(\ref{eq:trotter_2})  is replaced with an exponential, $e^{C \frac{\partial}{\partial P}}$. Such an approximation is  valid when the changes in environmental momenta are small \cite{kapral99}. 

Upon applying the MJ approximation, the nonadiabatic propagator can be written as
\begin{equation}\label{eq:MJA}
e^{-\mathcal{J}\delta} \approx Q_1(\delta)  e^{C \frac{\partial}{\partial P} } + \mathcal{O}(\delta^2) = \mathcal{M}(\delta) + \mathcal{O}(\delta^2),
\end{equation} where the explicit form of the MJ matrix $\mathcal{M}(\delta)$ for a two level system is \cite{mackernan08} \begin{widetext}
\begin{eqnarray}\label{eq:MJ_matrix}\mathcal{M}(\delta)=\left(\begin{array}{cccc}\cos^2{a} & -\hat{j}_{12} \cos{a}\sin{a}  & -\hat{j}_{12} \cos{a}\sin{a}  & \hat{j}_{1\rightarrow2}\sin^2{a} \\  \hat{j}_{21}\cos{a}\sin{a} & \cos^2{a} & -\sin^2{a}  & -\hat{j}_{12} \cos{a}\sin{a} \\ \hat{j}_{21}\cos{a}\sin{a}  &  -\sin^2{a} & \cos^2{a} & \hat{j}_{12} \cos{a}\sin{a} \\\hat{j}_{2\rightarrow1}\sin^2{a} & \hat{j}_{21} \cos{a}\sin{a} & \hat{j}_{21} \cos{a}\sin{a} & \cos^2{a}\end{array}\right).
\end{eqnarray} \end{widetext}

The MJ matrix $\mathcal{M}$ contains the operators $\hat{j}_{\alpha \beta}$ and $\hat{j}_{\alpha \rightarrow \beta}$ which give rise to the momentum changes in the bath that accompany the nonadiabatic transitions occurring in the quantum subsystem. The $\hat{j}_{\alpha \beta}$ operators are applied when the quantum subsystem makes a transition between diagonal and off-diagonal adiabatic surfaces, whereas $\hat{j}_{\alpha \rightarrow \beta}$ are applied when the subsystem makes a transition between two diagonal adiabatic surfaces. The action of these operators results in a change in the momentum of the bath, $\Delta P$, where
\begin{eqnarray}\label{eq:P_trans}
\nonumber \Delta P_{\alpha \beta}& =& \hat{d}_{\alpha \beta} {\rm sgn}(P\cdot\hat{d}_{\alpha \beta}) \sqrt{(P\cdot\hat{d}_{\alpha \beta})^2+M \hbar \omega_{\alpha \beta}} \\&&- (P\cdot\hat{d}_{\alpha \beta}), \nonumber  \\
\nonumber \Delta P_{\alpha \rightarrow \beta} &=& \hat{d}_{\alpha \beta} {\rm sgn}(P\cdot\hat{d}_{\alpha \beta}) \sqrt{(P\cdot\hat{d}_{\alpha \beta})^2+2 M \hbar \omega_{\alpha \beta}} \\ &&- (P\cdot\hat{d}_{\alpha \beta}).
\end{eqnarray} The resulting dynamics is a surface hopping scheme in which classical evolution on a single adiabatic surface is interspersed with quantum transitions accompanied by classical momentum shifts. 

There are two major differences between this MJ surface hopping scheme and the more commonly adopted FSSH approach. Firstly, in FSSH the bath dynamics is prescribed to only be performed on diagonal adiabatic surfaces. However, in the MJ approach the dynamics of the bath degrees of freedom can occur on either single adiabatic surfaces or the mean of two adiabatic surfaces, during which the system accumulates a dynamical phase (Eq.(\ref{eq:QCL_diag})). Secondly, the jump selection criteria in each approach are different as described in more detail below. 

Trajectories are generated by the action of  the short time propagators in Eq.(\ref{eq:total_prop}) which involves alternate applications of the adiabatic and nonadiabatic propagators. The adiabatic part of the QCL operator in Eq.(\ref{eq:QCL_diag}) can be written as the product of a phase factor, and a classical evolution propagator. The phase factor, $\mathcal{W}_{s_j} (t,t+\delta)$, associated with evolution on a surface $s_j$ from time $t$ to time $t+\delta$ is
\begin{equation}\label{eq:q_phase}
\mathcal{W}_{s_j} (t,t+\delta) = e^{i \int_{t}^{t+\delta} d\tau \omega_{s_j} (R_{s_j} (\tau))}, \end{equation} 
where $R_{s_j} (\tau)$ are the coordinates of the bath at time $\tau$ arising from classical evolution on state $s_j$. Hence,  the total short time propagator can be written as
\begin{eqnarray}\label{eq:short_t_prop}
\left(e^{i\mathcal{L}\delta}\right)_{s_{j-1} s_j} &\approx&  \mathcal{W}_{s_{j-1}} (t-\delta, t - \delta/2) (e^{iL\delta/2})_{s_{j-1}} \\\nonumber  && \times \mathcal{M}_{s_{j-1} s_j}(\delta) \mathcal{W}_{s_j} (t-\delta/2, t) (e^{iL\delta/2})_{s_{j}}.
\end{eqnarray} 
The nonadiabatic propagator is applied by Monte Carlo sampling from $\mathcal{M}(\delta)$ \cite{mackernan08}. If the system is initially in state $s_{j-1}$, a new state index $s_j$ is chosen by sampling from the set of possible system states with the following probability 
\begin{equation}\label{eq:P_jump}
P(s_j | s_{j-1}, \delta) = \frac{|(Q_1)_{s_{j-1} s_j}(\delta)| }{ \sum_{s_j} |(Q_1)_{s_{j-1} s_j}(\delta)|}.
\end{equation}
If the system has sufficient momentum along the direction of the nonadiabatic coupling vector to access the new state, then the jump is accepted and the momentum shift is performed. The trajectory is then weighted by the inverse of the jumping probability given in Eq.(\ref{eq:P_jump}). By virtue of this sampling procedure, each time a jump is sampled the trajectory is re-weighted by the factor $\sum_{s_j} |(Q_1)_{s_{j-1} s_j}(\delta)|$. It is the multiplicative accumulation of these weights, and the associated phase factors, which leads to numerical difficulties with the convergence of this method at long times \cite{mackernan08}. 

Hence to generate the time-evolution of an observable, after sampling from the initial density, trajectories are propagated in the following manner:
\begin{enumerate}
  \item  The bath degrees of freedom are evolved classically from an initial phase point $(R,P)$ to $(R',P')$, on the $s_{j-1}$ adiabatic surface through a time-segment of length $\delta/2$.
  \item  A new subsystem state index $s_j$ is determined by Monte Carlo sampling using the probability measure given in Eq.(\ref{eq:P_jump}).
  \item  The transition to the state $s_j$ is assessed; if the momentum shift to be applied is imaginary then the transition is forbidden and the jump is rejected. If the transition is allowed, the momentum shift (if any) is applied, such that $(R',P')\rightarrow(R',P'+\Delta P)$, where $\Delta P$ is given by Eq.(\ref{eq:P_trans}), and the subsystem state label is changed from $s_{j-1}$ to $s_j$. If the jump is rejected, then $s_j = s_{j-1}$.
  \item  The statistical weight of the trajectory is multiplied by the denominator of Eq.(\ref{eq:P_jump}).
  \item  The bath degrees of freedom are evolved classically on the $s_{j}$ adiabatic surface through a time-segment of length $\delta/2$.  
\end{enumerate}
This process is then repeated until the required number of time-steps have been performed. 

\subsection{The Generalized Quantum Master Equation}
\label{subsec:GQME}
Although the MJ surface hopping approach provides an accurate method to propagate nonadiabatic quantum dynamics, as we will see in Sec.(\ref{sec:results}), the rapid accumulation of statistical weights makes the direct application of this scheme intractable in many cases. In this section we show how the MJ scheme can be combined with the GQME formalism to yield an approach which allows for the generation of long time nonadiabatic dynamics.

As discussed in Sec.(\ref{sec:intro}), in many condensed phase systems one is primarily interested in the time-evolution of the reduced density of the subsystem, which is defined as
\begin{equation}\label{eq:rdm}
\hat{\rho}_{s} (t) = Tr_b ( \hat{\rho} (t) ) = \int dX \hat{\rho}_W (X,t),
\end{equation}
where $\hat{\rho} (t)$ is the density operator for the entire system, and $Tr_b$ indicates the partial trace taken 
over the bath degrees of freedom. In the problem that we consider in Sec.(\ref{sec:results}) the initial state of the system can be factorized in the following manner,
\begin{equation}\label{eq:ini_con}
\hat{\rho}(t=0) = \hat{\rho}_{s}(0) \otimes \hat{\rho}^{eq}_{b},
\end{equation}
where 
\begin{equation}\label{eq:bath_density}
\hat{\rho}^{eq}_{b} = \frac{\exp(-\beta \hat{H}_b)}{Tr_b [ \exp(-\beta \hat{H}_b) ] }
\end{equation} 
is the density operator for the isolated bath in thermal equilibrium. In what follows we will assume such a factorization however this condition need not be applied in general \cite{geva06}. We assume that the coupling part of the Hamiltonian, $\hat{H}_{sb}$ is of the form,
\begin{equation}\label{eq:H_sb}
\hat{H}_{sb} = \hat{S} \otimes \hat{\Lambda}, 
\end{equation}
where $\hat{S}$ is a pure subsystem operator, and $\hat{\Lambda}$ is a pure bath operator. We will also write the bath part of the coupling Hamiltonian, $\hat{\Lambda}$, such that its thermal average vanishes,
\begin{equation}\label{eq:bath_avg}
\langle \hat{\Lambda} \rangle_{eq} = Tr_b [ \hat{\Lambda} \hat{\rho}^{eq}_b ]  = 0,
\end{equation} 
where $\langle \cdots \rangle_{eq} = Tr_b \left( \cdots \hat{\rho}^{eq}_b \right)$ denotes the equilibrium bath average. In the problem we consider in Sec.~\ref{sec:results} such a condition is naturally satisfied however in general cases it can be simply incorporated by redefining $\hat{\Lambda}$ relative to its thermal average \cite{shi03a}.

Under these conditions, the time evolution of the subsystem RDM is given by the formally exact Nakajima-Zwanzig GQME \cite{nakajima58,zwanzig60},
\begin{equation} \label{eq:GQME}
\frac{d}{dt} \hat{\rho}_{s} (t) = -i\mathcal{L}_{s} \hat{\rho}_s (t) - \int_0^{t} d\tau \mathcal{K}(\tau) \hat{\rho}_s(t-\tau)
\end{equation} 
where the subsystem Liouville operator is $\mathcal{L}_s = \frac{1}{\hbar}[ \hat{H}_s, \cdot]$. The memory kernel is given by, 
\begin{equation} \label{eq:kern}
\mathcal{K}(\tau)  = Tr_b \{ \mathcal{L}_{sb} \exp(-i \mathcal{QL}\tau)\mathcal{QL}_{sb} \hat{\rho}^{eq}_{b} \}
\end{equation}
where $\mathcal{P}$ and $\mathcal{Q} $ are the projection operators, $\mathcal{P} = \hat{\rho}_b^{eq}\otimes Tr_b (\cdot)$, and $\mathcal{Q} = 1-\mathcal{P}$. The free subsystem evolution prescribed by $\mathcal{L}_s$ is generally very simple to simulate, and hence in the GQME picture, calculating the evolution of the subsystem RDM reduces to the calculation of the memory kernel, $\mathcal{K}(\tau)$.

The general form for the memory kernel, given above, is not straightforward to evaluate since it explicitly depends  
on the projection operator, $\mathcal{P}$.  An elegant solution to this problem was presented by Shi and Geva \cite{shi03a} and involves rewriting the memory kernel using the following relation
\begin{eqnarray} \label{eq:trick}
&&e^{-i (\mathcal{L}-\mathcal{L}_{sb} \mathcal{P} ) \tau} =  e^{-i \mathcal{L} \tau} \\\nonumber
&& + i\int_0^{\tau}d\tau' e^{-i \mathcal{L} (\tau-\tau')} \mathcal{L}_{sb}\mathcal{P}e^{-i (\mathcal{L}-\mathcal{L}_{sb}\mathcal{P})\tau'}.
\end{eqnarray} 
Upon inserting this relation into (\ref{eq:kern}), one finds
\begin{eqnarray} \label{eq:k_tot}
\mathcal{K}(\tau) & = & \mathcal{K}_1(\tau) + i \int_0^{\tau}d\tau' \mathcal{K}_1(\tau - \tau') \mathcal{K}_2(\tau),
\end{eqnarray} where \begin{eqnarray} \label{eq:k_1}
\mathcal{K}_1(\tau) & = & Tr_b \{ \mathcal{L}_{sb} e^{-i \mathcal{L}\tau}\mathcal{L}_{sb} \hat{\rho}^{eq}_{b} \}, 
\end{eqnarray} and \begin{eqnarray} \label{eq:k_2a}
\mathcal{K}_2(\tau) & = & Tr_b \{ e^{-i (\mathcal{L}-\mathcal{L}_{sb} \mathcal{P} ) \tau} \hat{\rho}^{eq}_{b} \}. 
\end{eqnarray}
Another insertion of Eq.(\ref{eq:trick}) into the Eq.(\ref{eq:k_2a}) above yields
\begin{eqnarray} \label{eq:k_2b}
\mathcal{K}_2(\tau) & = & \mathcal{K}_3(\tau) + i \int_0^{\tau}d\tau' \mathcal{K}_3(\tau - \tau') \mathcal{K}_2(\tau), 
\end{eqnarray} where \begin{eqnarray} \label{eq:k_3} 
\mathcal{K}_3(\tau) & = & Tr_b \{ e^{-i \mathcal{L}\tau}\mathcal{L}_{sb} \hat{\rho}^{eq}_{b} \}.
\end{eqnarray}
Using this approach the full memory kernel, $\mathcal{K}$, required to propagate the subsystem RDM, can be constructed via direct simulation of the unprojected dynamics. From these simulations one can generate the partial memory kernels $\mathcal{K}_1$ and $\mathcal{K}_3$ which can then be used to obtain $\mathcal{K}$, by solving Eqs. (\ref{eq:k_2b}) and (\ref{eq:k_tot}). 
 
The matrix elements of  $\mathcal{K}_1$ and $\mathcal{K}_3$, in any basis which spans the subsystem Hilbert space, are given by
\begin{eqnarray} \label{eq:k_1_elements}
\nonumber(\mathcal{K}_1)_{\alpha \alpha' \beta \beta' }(\tau) &=& \Big\langle S_{\alpha \mu'}(\tau) \hat{\Lambda}^{\beta' \alpha'}_{\mu' \mu} (\tau) S_{\mu \beta}(0)  \hat{\Lambda}(0)  \Big\rangle_{eq} \\\nonumber &&-  \Big\langle S_{\mu' \alpha'}(\tau) \hat{\Lambda}^{\beta' \mu'}_{\alpha  \mu} (\tau) S_{\mu \beta}(0)  \hat{\Lambda}(0) \Big\rangle_{eq} \\\nonumber &&+  \Big\langle \hat{\Lambda}(0) S_{\beta' \mu}(0)   \hat{\Lambda}^{\mu \mu'}_{\alpha  \beta} (\tau) S_{\mu' \alpha'}(\tau) \Big\rangle_{eq} \\\nonumber && -\Big\langle  \hat{\Lambda}(0) S_{\beta' \mu}(0) \hat{\Lambda}^{\mu \alpha'}_{\mu'  \beta}(\tau) S_{\alpha \mu'}(\tau)  \Big \rangle_{eq},\\
\end{eqnarray} 
\begin{eqnarray}\label{eq:k_3_elements}
\nonumber (\mathcal{K}_3)_{\alpha \alpha' \beta \beta' }(\tau) &=& \Big\langle (\hat{1}_b)^{\beta' \alpha'}_{\alpha  \mu}(\tau) S_{\mu \beta}(0) \hat{\Lambda}(0)  \Big\rangle_{eq} \\ &&-  \Big\langle S_{\beta' \mu}(0)  \hat{\Lambda}(0) (\hat{1}_b)^{\mu \alpha'}_{\alpha  \beta}(\tau)  \Big\rangle_{eq},
\end{eqnarray} 
where $\hat{1}_b$ is the unit operator for the bath. In the above two expressions the Einstein summation convention is used.

The above expressions for the matrix elements of the partial memory kernels, $\mathcal{K}_1$ and $\mathcal{K}_3$, contain correlation functions of the following form, 
\begin{eqnarray} \label{eq:sdbcf1}
\langle \hat{\Lambda} \hat{\Gamma}^{\beta' \alpha'}_{ \alpha  \beta}(\tau) \rangle_{eq} =Tr \left( | \beta \rangle \hat{\rho}_b^{eq} \hat{\Lambda}  \langle \beta' |  e^{i \mathcal{L}\tau/\hbar} | \alpha' \rangle  \hat{\Gamma} \langle \alpha | \right),
\end{eqnarray}  
where we have used $\hat{\Gamma}$ to denote a general bath operator, which in the cases outlined above is either $\hat{1}_b$ or $\hat{\Lambda}$. 

If we define two new operators, $\hat{A} = \hat{\Lambda}\otimes |\beta\rangle\langle\beta'|$ and $\hat{B}=\hat{\Gamma}\otimes |\alpha'\rangle\langle\alpha|$, which are operators on the full system, then expression (\ref{eq:sdbcf1}) takes on the general form for a quantum time correlation function, 
\begin{eqnarray} \label{eq:sdbcf2}
\langle \hat{\Lambda} \hat{\Gamma}^{\beta' \alpha'}_{ \alpha  \beta}(\tau) \rangle_{eq} =  Tr(\hat{\rho}_b^{eq}\hat{A} \hat{B}(\tau)).
\end{eqnarray}
with the difference that the equilibrium density corresponds to that of the entire system, while in the latter case it refers to that of the isolated bath. 

An approximate solution to Eq.(\ref{eq:sdbcf2}) can then be constructed using the QCLE in the adiabatic basis. Working in the coordinate representation of the bath degrees of freedom, and making use of the partial Wigner transform Eq.(\ref{eq:sdbcf2}) can be rewritten as
\begin{eqnarray} \label{eq:c_ab}
 Tr(\hat{\rho}_b^{eq}\hat{A} \hat{B}(\tau)) = Tr_s \int dX \left[\hat{\rho}_b^{eq} \hat{A}\right]_W(X,0) \hat{B}_W(X,\tau),
\end{eqnarray} 
The time evolution of $\hat{B}_W$ can be approximated using QCL evolution in the adiabatic basis, as described in the previous section, and this information can then be used to form the partial memory kernels in the subsystem basis. The memory kernel can then be constructed, and the subsystem RDM propagated as follows:
\begin{enumerate}
\item The MJ algorithm described in Sec.(\ref{subsec:MJ})  is used to obtain the correlation functions necessary to form $\mathcal{K}_1$ and $\mathcal{K}_3$. 
 \item $\mathcal{K}_2$ is generated from $\mathcal{K}_3$ by an iterative solution to Eq.(\ref{eq:k_2b}), using $\mathcal{K}_3$ itself as an initial guess for $\mathcal{K}_2$. This iterative procedure typically converges very quickly, and often requires only a few tens of iterations. 
 \item $\mathcal{K}_1$ and $\mathcal{K}_2$ are used as input to obtain the full memory kernel $\mathcal{K}$ by numerical integration of Eq.(\ref{eq:k_1}). 
 \item  Using the full memory kernel, the evolution of the subsystem density is generated by direct numerical integration of the GQME using Eq.~(\ref{eq:GQME}).
 \end{enumerate}
Using this approach one can propagate the subsystem RDM for long times using only short-time information obtained from MJ trajectories performed in the adiabatic basis. This forms the basis of the MJ-GQME approach for which we assess the accuracy and efficiency in the following section.

\section{Results and Discussion} \label{sec:results}

In order to assess the accuracy and efficiency of our MJ-GQME approach, we performed simulations of the spin-boson model. Despite its apparent simplicity, this system is a prototypical model for the study quantum transport and relaxation processes in the condensed phase \cite{leggett87,weiss92}, and remains a challenging test to approximate methods. Although the QCLE is formally exact for the spin-boson model, the MJ approximation has been invoked and hence the MJ-GQME technique is not guaranteed to be exact. Since a wealth of numerically exact results are available for various parameter regimes of the spin-boson model, it provides an ideal benchmark test case for the accuracy and efficiency of approximate nonadiabatic dynamics approaches.  

\subsection{Spin-Boson Model} \label{subsec:model}
The spin-boson Hamiltonian can be written in the subsystem basis as 
\begin{equation} \label{eq:sb_ham}
\hat{H} = \epsilon \hat{\sigma}_z + \Delta \hat{\sigma}_x + \frac{\hat{P}^2}{2M} + \sum_j\left(\frac{1}{2}M_j\omega_j^2 \hat{R}_j^2 - c_j \hat{R}_j\hat{\sigma}_z\right).
\end{equation} This Hamiltonian describes a two level quantum system with energetic bias $2\epsilon$, and electronic coupling (or tunneling matrix element) $\Delta$, that is bi-linearly coupled to a bath of independent harmonic oscillators. In this model the interaction between the system and the bath can be fully characterized by the spectral density, $J(\omega)$, which we will choose to be of the Ohmic form,
\begin{eqnarray} \label{eq:spec_dens}
J(\omega) & = & \frac{\pi}{2} \xi \omega e^{-\omega / \omega_c}. 
\end{eqnarray} The Kondo parameter $\xi$ controls the strength of the coupling between subsystem and the bath, and the cutoff frequency $\omega_c$ sets the primary time-scale for the bath evolution. Here we invoke initial conditions where the subsystem population all starts in diabatic state 1 and the bath positions and momenta correspond to their equilibrium distribution in isolation. These conditions correspond to situations where the initial preparation of the subsystem occurs quickly on the time-scale of the bath relaxation, which is typically the case in photoexcited processes. 

\begin{figure} 
\includegraphics[width=\columnwidth,angle=0]{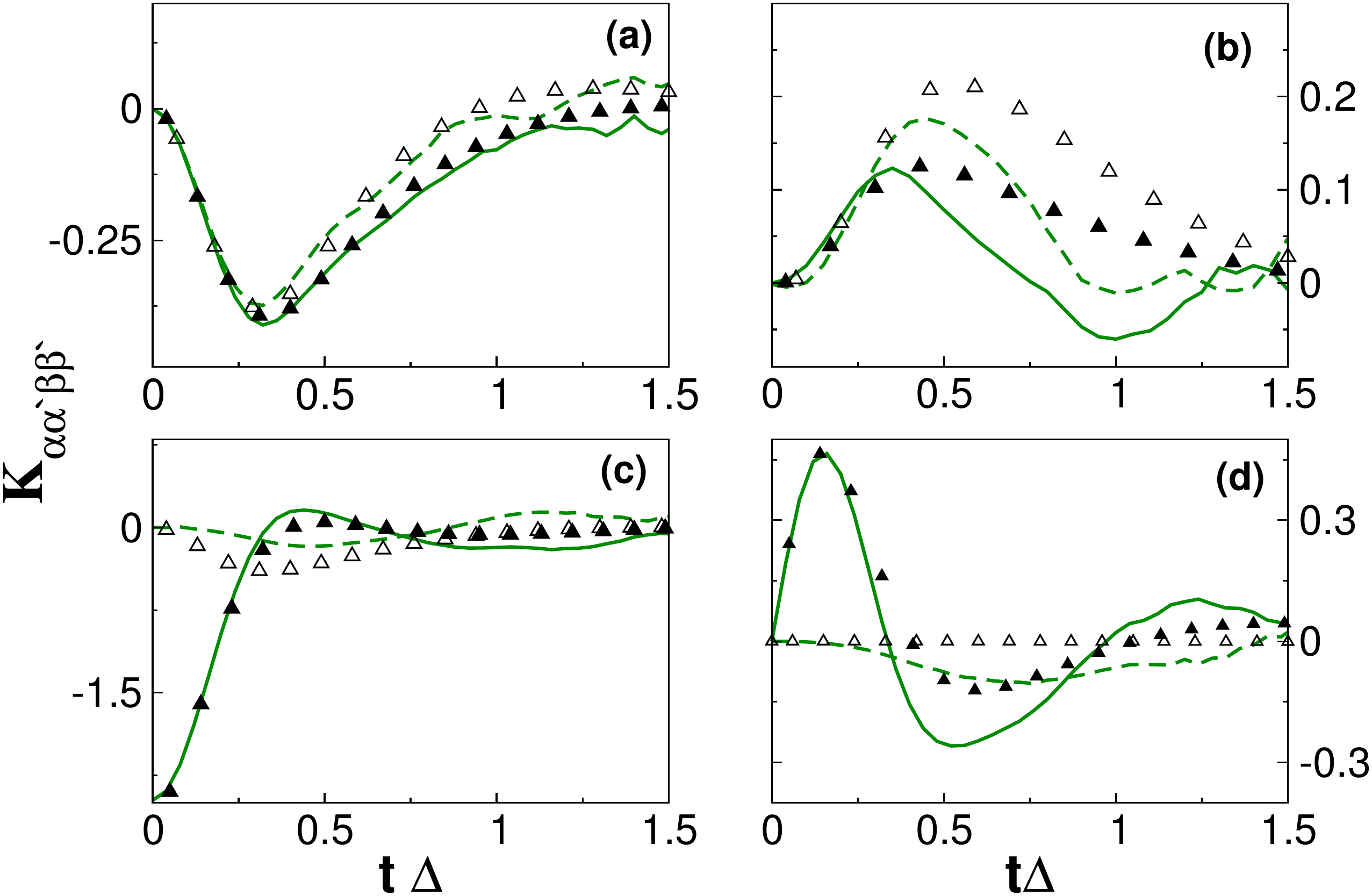} 
\caption{Matrix elements of the memory kernel of the GQME for for $\omega_c= 2.5 \Delta$, $\epsilon = \Delta$, $\beta = 5 \Delta^{-1}$, $\xi=0.2$, $\delta = 0.02\Delta^{-1}$, and $N_{traj} =2$x$10^6$.  (a) Real and (b) imaginary  parts of $\mathcal{K}_{1211}$ (solid line, filled triangles) and $\mathcal{K}_{1222}$ (dashed line, open triangles). (c) Real and (d) imaginary parts of $\mathcal{K}_{1212}$ (solid line, filled triangles) and $\mathcal{K}_{1221}$ (dashed line, open triangles).  In each panel the MJ based results are given in solid and dashed green lines, and the exact QUAPI results are given in filled and open black triangles.}
\label{fig:1}
\end{figure}
 
The Wigner transform of the equilibrium density for the isolated bath is
\begin{eqnarray}\label{eq:rho_b_eq}
\nonumber \rho_{b,W}^{eq}(X) &= &\prod_{j}\frac{\tanh(\beta \omega_j/2)}{\pi} \\\nonumber && \times \exp\left[-\frac{2 \tanh(\beta \omega_j/2)}{\omega_j}\left(\frac{P_j^2}{2M_j} + \frac{M_j\omega_j^2}{2}R_j^2 \right)\right], \\
\end{eqnarray} 
which is used for sampling the bath initial conditions. The Wigner transform of the following quantities, $(\hat{\Lambda}\hat{\rho}_b^{eq})_W$ and $(\hat{\rho}_b^{eq}\hat{\Lambda})_W$, are also required. In this particular case $\hat{\Lambda} = c_j \hat{R}_j$ and hence the Wigner transforms can be performed analytically \cite{imre67}, 
\begin{equation} \left(\hat{\Lambda}\hat{\rho}_b^{eq}\right)_W =\left(\hat{\rho}_b^{eq}\hat{\Lambda}\right)_W^* = \Lambda_W\rho_{b,W}^{eq} - \frac{i}{2\hbar}\frac{\partial \Lambda_W}{\partial R}\cdot\frac{\partial \rho_{b,W}^{eq}}{\partial P}. 
\end{equation}
In our calculations 200 bath modes were used to represent the continuous spectral density which, for all regimes and approaches employed, gave results converged to graphical accuracy.

\begin{figure} 
\includegraphics[width=0.85\columnwidth,angle=0]{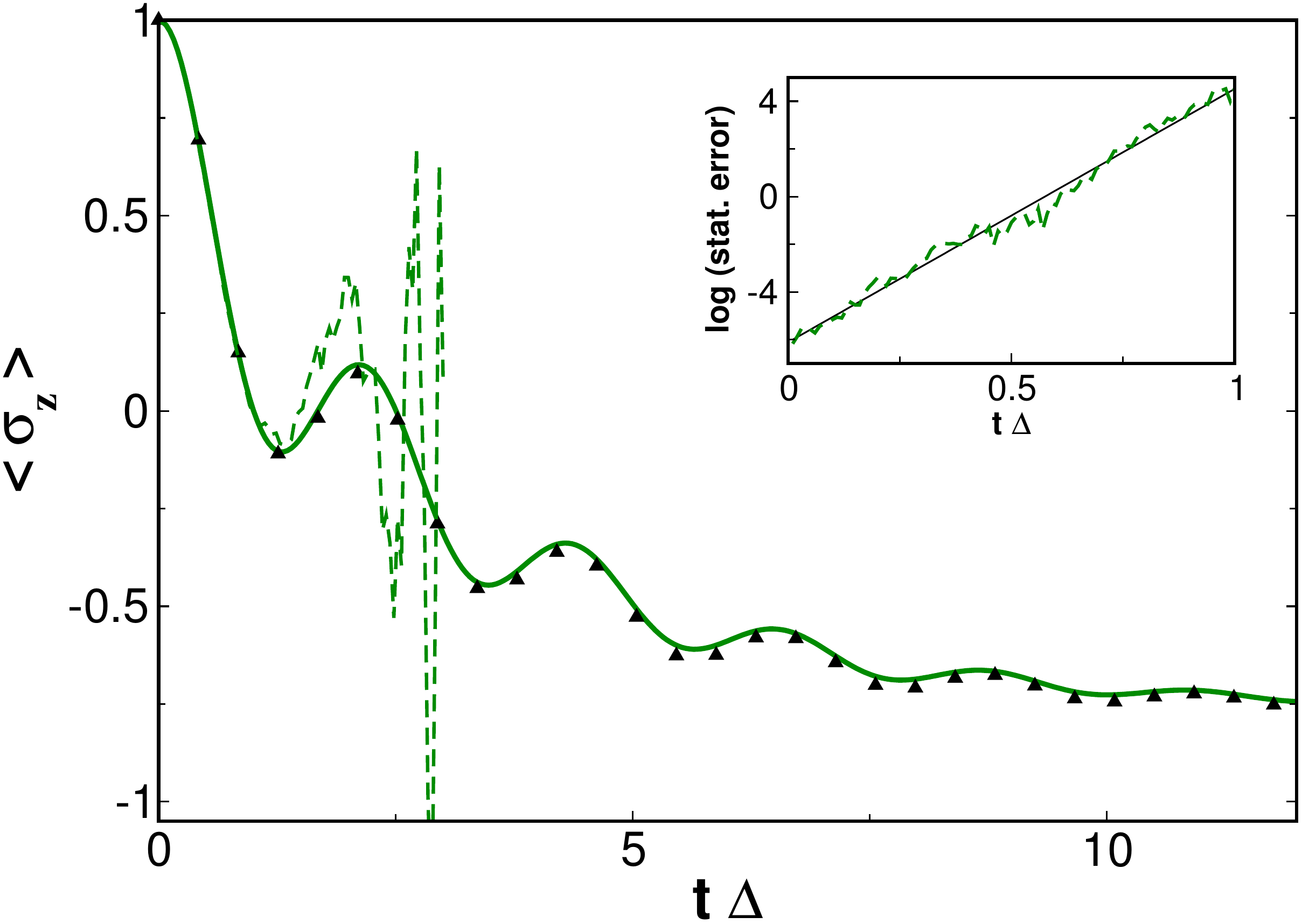} 
\caption{Population dynamics for $\omega_c= 2.5 \Delta$, $\epsilon = \Delta$, $\beta = 5 \Delta^{-1}$, $\xi=0.2$, $\delta = 0.02\Delta^{-1}$, and $N_{traj} =2$x$10^6$. Inset: Semi-log plot showing the exponential growth of the statistical error of the direct MJ-QCL approach as a function of time.  Direct MJ results (dashed green line), MJ-GQME (solid green line), and exact QUAPI results (filled black triangles).}
\label{fig:2}
\end{figure}

The memory kernel, $\mathcal{K}$, for the spin-boson model has four independent non-zero elements in the subsystem basis. Fig. \ref{fig:1} compares the memory kernel elements obtained from the MJ-GQME with numerically exact results calculated using the QUAPI method \cite{makri94a,makri94b,shi03a}. The matrix elements are reproduced quite well by the MJ-GQME method, confirming the viability of the the MJ approximation. However, despite the small deviations from the exact memory kernel Fig. \ref{fig:2} shows that using the MJ elements in the GQME formalism leads to exact agreement with the exact population dynamics. However a direct application of the MJ algorithm using the same number of trajectories deviates markedly from the exact results due to the rapid increase in the statistical error with propagation time which (as demonstrated in the inset of Fig. \ref{fig:2}) scales exponentially with time. Our MJ-GQME method allows long time dynamics to be obtained by instead calculating the memory kernel elements and then using these to propagate the subsystem RDM. In Fig. \ref{fig:1} it can be observed that all of these kernel elements decay on timescales on the order of 1.5$\Delta^{-1}$. Hence by only performing short-time trajectories the entire population decay, which occurs in a time of 15$\Delta^{-1}$, can be obtained. Accessing this time-scale using the direct MJ approach, which has exponential growth in the statistical error, would require many orders of magnitude more trajectories in order to converge on that time-scale.

The ability of our MJ-GQME approach to obtain accurate population dynamics in nonadiabatic regimes using only short trajectories as input is highly promising since these regimes provide a great challenge to other approximate approaches which are applicable to complex systems. Some of the most commonly used approaches are FSSH \cite{tully90}, Ehrenfest dynamics \cite{mclachlan64}, and linearized approaches such as the Poisson-bracket mapping equation (PBME) \cite{kim-map08}. The PBME is very closely related to the linearized semiclassical initial-value representation (LSC-IVR)\cite{sun98} and other linearized approaches \cite{huo12}, and in previous studies these theories have been seen to give similar results \cite{ananth07, tao10, kelly11, kelly12}. 

In this model, where the both the initial condition and the observable of interest are specified in the diabatic basis, the implementation of the FSSH algorithm requires some care. This is because the transformation between the basis representations in FSSH neglects evolution on off-diagonal surfaces. As a result, one has no knowledge of the off-diagonal elements of the density matrix (the so-called coherence problem) and hence converting the adiabatic populations, produced by the FSSH evolution, to diabatic populations is ill defined except in regions where the electronic coupling vanishes. In the spin boson model the electronic coupling does not vanish anywhere and so care must be taken in initializing the FSSH trajectories, and in obtaining the diabatic state population difference as a function of time. Hence we use the prescription of M\"uller and Stock to collapse trajectories onto adiabatic surfaces in a way consistent with the diabatic initial condition, and to extract the diabatic populations by performing a basis transform using the adiabatic coefficients \cite{stock97}.

\begin{figure} 
\includegraphics[width=0.8\columnwidth,angle=0]{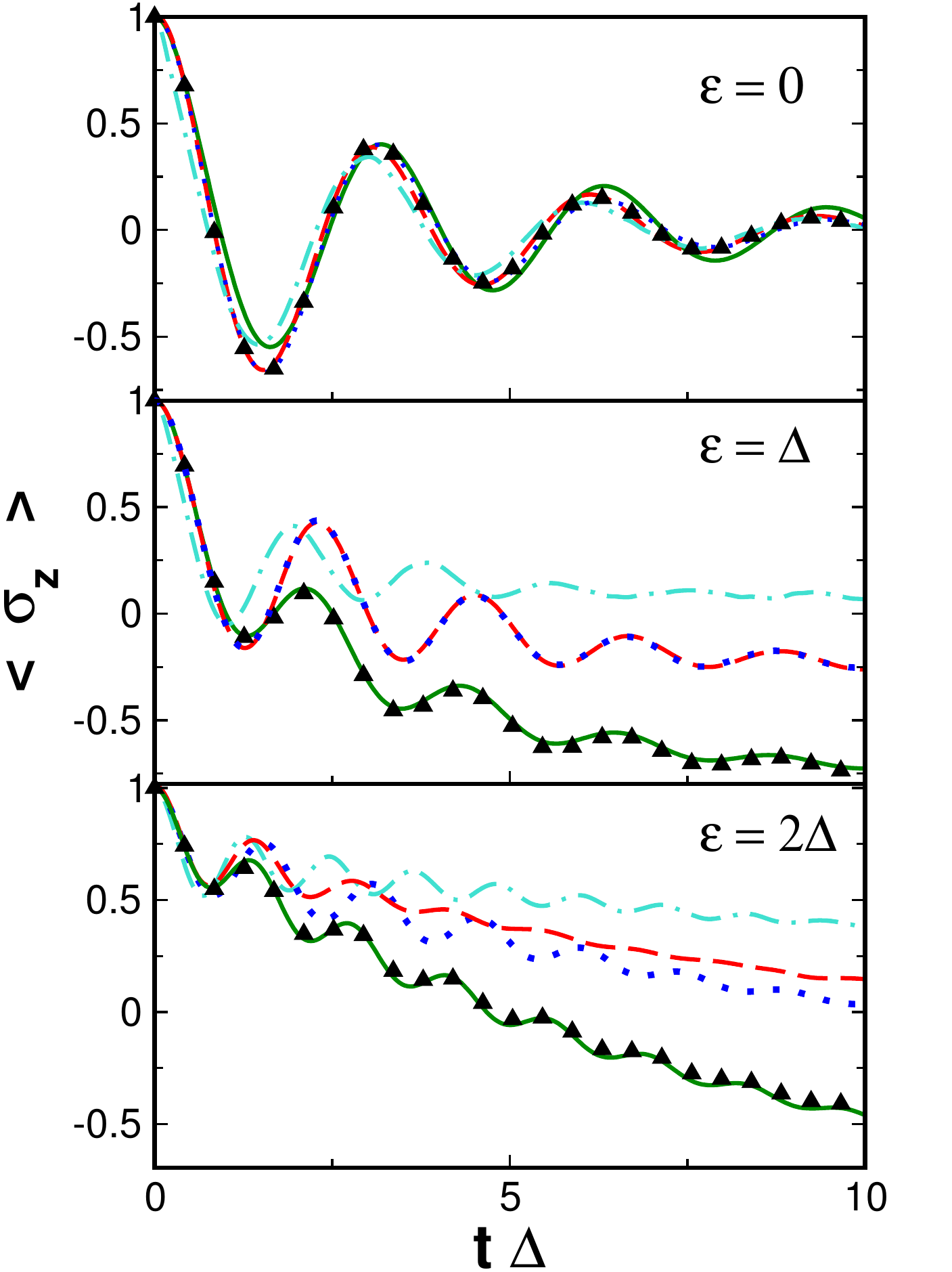} 
 \caption{Effect of changing the electronic bias on the population dynamics, with $\omega_c= 2.5 \Delta$, and $\beta = 5 \Delta^{-1}$, and $\xi=0.2$.  Exact QUAPI results (black filled triangles), MJ-GQME (green full line), PBME (red dashed line), Ehrenfest (navy dotted line), and FSSH (turquoise dot-dashed line).} 
\label{fig:3}
\end{figure}

Figure \ref{fig:3} shows the population dynamics for a nonadiabatic regime, ($\omega_{c}/\Delta>1$) as the energetic bias between the two levels is increased. When the electronic levels are degenerate ($\epsilon=0$)  all the schemes are observed to give quantitative agreement with the phase and amplitude of the population dynamics at all times. However, upon increasing the energetic bias (middle and lower panels) FSSH, Ehrenfest, and PBME all fail to reach the correct long-time populations of the states. This failure is due to an incorrect treatment of the exchange of energy between the quantum subsystem and the bath, giving rise to a violation of detailed balance, which results in populations which are too close to being equal, $\langle\sigma_{z}\rangle=0$. Despite not relaxing to the correct populations the Ehrenfest and PBME accurately obtain reasonable agreement with the phase and amplitude of the population oscillations whereas FSSH accrues a phase shift as time progresses. In contrast the MJ-GQME results are in perfect agreement with the exact results.

\begin{figure}   
\includegraphics[width=0.8\columnwidth,angle=0]{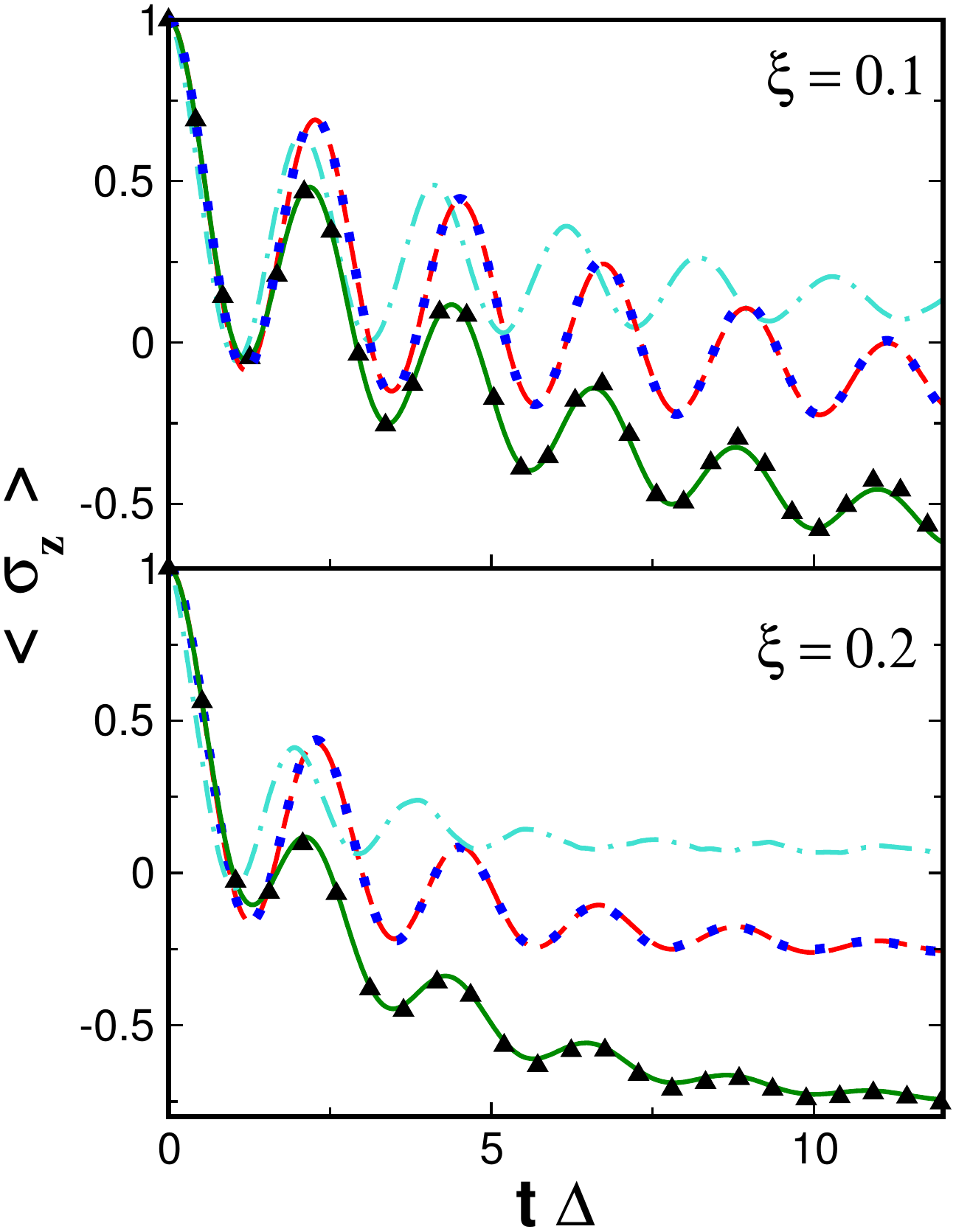}
 \caption{Effect of changing the system-bath coupling on the population dynamics,  with $\epsilon=\Delta$, $\omega_c=2.5 \Delta$, and $\beta = 5 \Delta^{-1}$. Exact QUAPI results (black filled triangles), MJ-GQME (green full line), PBME (red dashed line), Ehrenfest (navy dotted line), and FSSH (turquoise dot-dashed line).} 
\label{fig:4}
\end{figure}

The failures of most trajectory based methods are further exacerbated by increasing the coupling between the quantum subsystem and the bath. Figure \ref{fig:4} shows the population  dynamics of a biased system in the nonadiabatic regime, as the system-bath coupling is increased. In both cases Ehrenfest theory and the PBME are coincident, obtaining the correct phase and amplitude but failing to obtain the correct long time populations which becomes increasingly apparent as the coupling is increased. The FSSH results show a larger error at long times and exhibit a slight phase shift as time progresses \cite{martinez07}. As the coupling to the bath increases FSSH, PBME and Ehrenfest theory deviate from the exact results at progressively shorter times due to an incorrect treatment of the back-reaction of the bath. The importance of the back-reaction term increases as the system-bath coupling increases making the error more pronounced. Meanwhile the MJ-GQME approach gives the exact results.

\begin{figure}  
\includegraphics[width=0.8\columnwidth,angle=0]{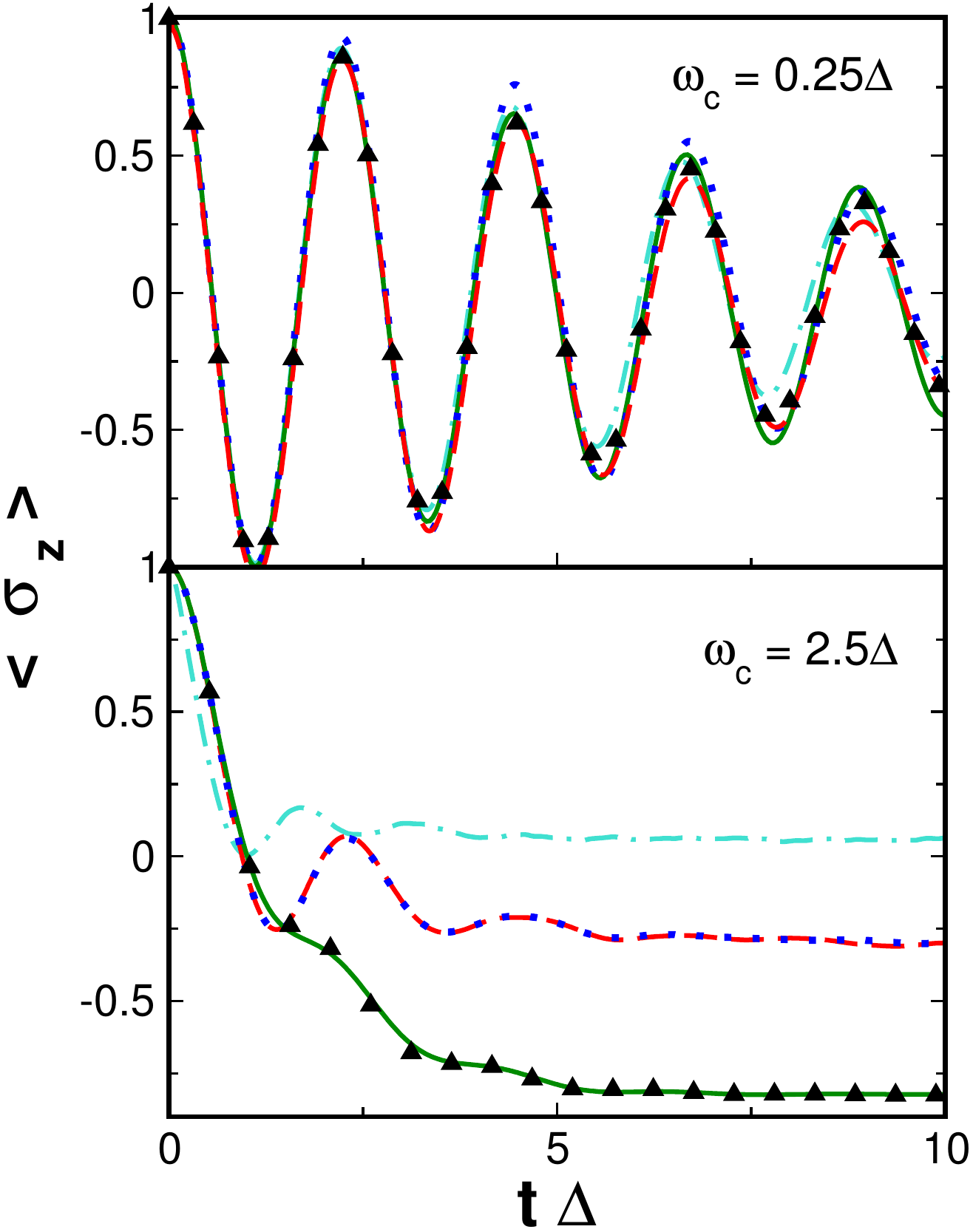}
\caption{Effect of changing the nonadiabaticity of the system on the population dynamics, with $\epsilon=\Delta$, $\beta = 5 \Delta^{-1}$, and $\xi=0.4$. Exact QUAPI results (black filled triangles), MJ-GQME (green full line), PBME (red dashed line), Ehrenfest (navy dotted line), and FSSH (turquoise dot-dashed line).}
\label{fig:5}
\end{figure}

Even in biased and strongly coupled regimes most trajectory based approaches are capable of yielding accurate results in the adiabatic regime ($\omega_c/\Delta<1$). Figure \ref{fig:5} depicts subsystem evolution as the nonadiabaticity, $\omega_c/\Delta$, is increased. In the adiabatic case ($\omega_c/\Delta = 0.25$), all the approaches are in good agreement with that exact results at short times, while FSSH is mildly underdamped at long times. In the nonadiabatic regime ($\omega_c/\Delta = 2.5$),  however FSSH and the mean field theories fail to correctly describe the relaxation of the system. Again, quantitative agreement with exact results is obtained using the MJ-GQME approach in both cases.

Since the spin-boson model is the prototypical model for charge transfer in solution, it is interesting to also investigate the Marcus regime of nonadiabatic electron transfer, corresponding to strong system-bath coupling and a high temperature bath. Our previous discussion has highlighted that strong system-bath coupling provides a challenge to traditional theories that is exacerbated further in nonadiabatic and/or biased regimes. As such this provides both a stringent test of our MJ-GQME approach, as well as the MJ approximation itself. The Marcus-Zusman expression for the thermal electron transfer rate constant is given by \cite{Marcus64,Marcus85,Zus80,Garg85}, 
\begin{equation}\label{eq:marcus}
k_{th} = \frac{\Delta^2}{1+\frac{2\pi\Delta^2}{\lambda\omega_c}}\sqrt{\frac{\pi\beta}{\lambda}}e^{-\beta\frac{(2\epsilon-\lambda)^2}{4\lambda}}\left(1+e^{-2\beta\epsilon}\right),
\end{equation} 
where $\lambda$ is the reorganization energy, which is $2 \xi \omega_{c}$ for the Ohmic spectral density. The rate can extracted from an exponential fit to the population dynamics, after an initial transient time. 

\begin{figure} 
\includegraphics[width=0.9\columnwidth,angle=0]{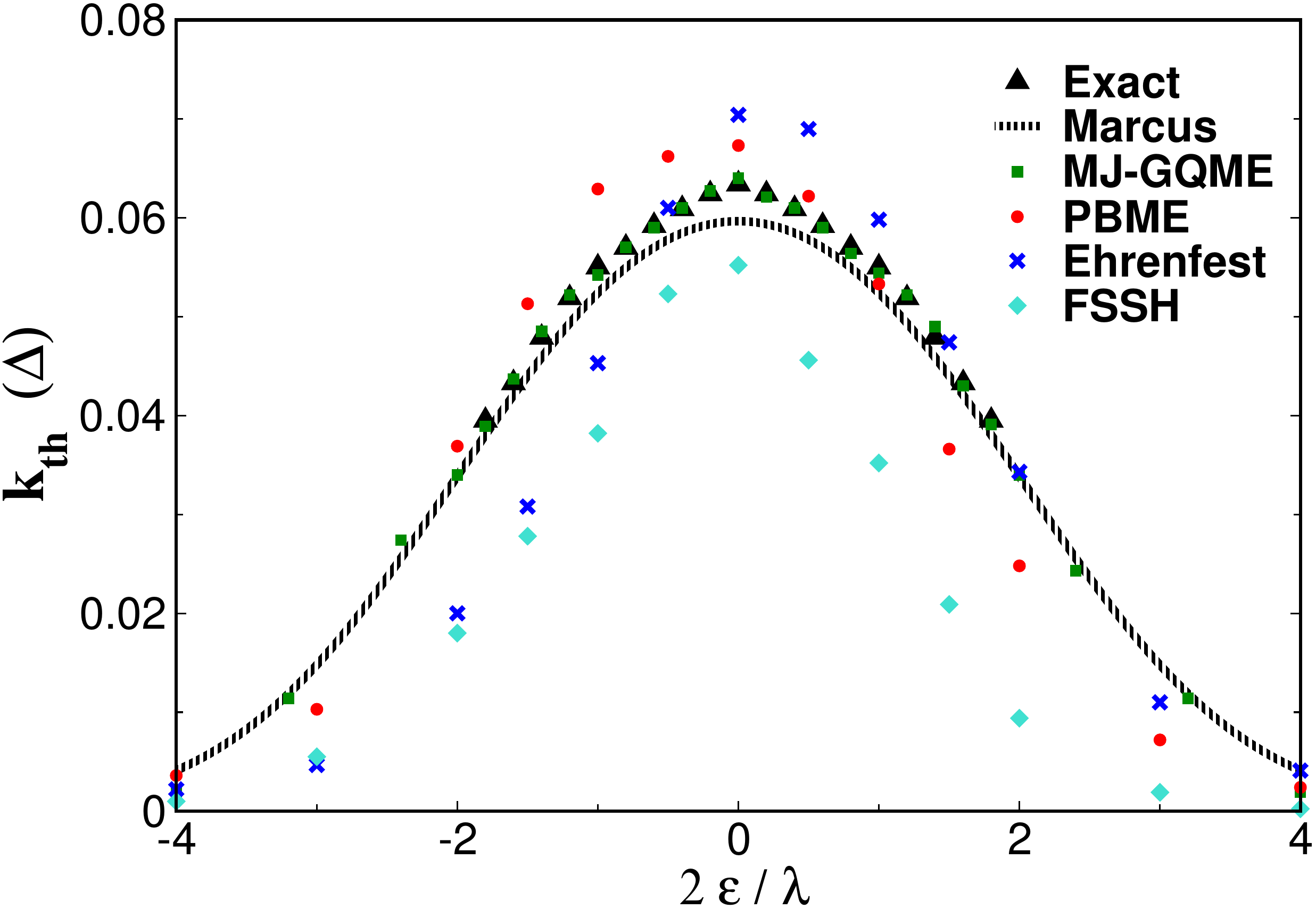}
\caption{Plot of the thermal electron transfer rate constant as a function of the driving force. The bath has an Ohmic spectral density and $\omega_c=2\Delta$, and $\beta = 0.05 \Delta^{-1}$, $\xi = 5$, and $\lambda = 20 \Delta$ is the reorganization energy. The exact QMC results are taken from reference \cite{Muhlbacher04} (black filled triangles), Marcus rate (broken black line), MJ-GQME (green line), PBME (filled red circles), Ehrenfest (navy crosses), and FSSH (turquoise filled diamonds).}
\label{fig:6}
\end{figure}

Figure \ref{fig:6} shows the rates compared to those obtained from Marcus theory and exact quantum Monte Carlo (QMC) results \cite{Muhlbacher04}. In this regime the Marcus rate slightly underestimates the exact QMC result at zero bias. Our MJ-GQME approach exactly captures the QMC data in the Marcus turnover regime for the thermal transfer rate as a function of the driving force, despite being in a nonadiabatic regime with strong system-bath coupling. The FSSH, Ehrenfest, and PBME results also reproduce the qualitative turnover behavior of the thermal transfer rate, however they are not in quantitative agreement with the exact results. 

\begin{figure}  
\includegraphics[width=0.9\columnwidth,angle=0]{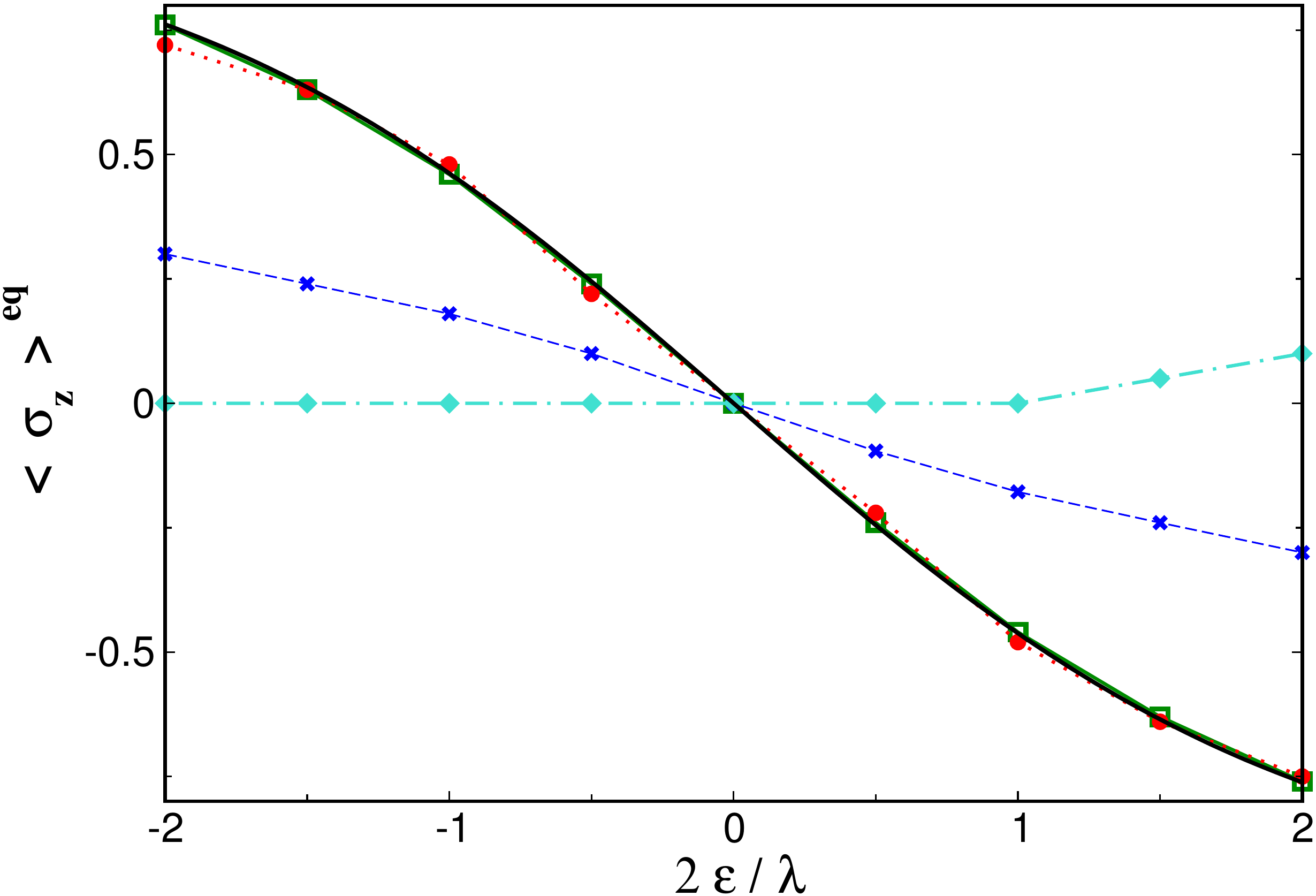}
\caption{Plot of the equilibrium population difference for the electronic system as a function of the driving force. The bath has an Ohmic spectral density and $\omega_c=2\Delta$, and $\beta = 0.05 \Delta^{-1}$, $\xi = 5$, and $\lambda = 20 \Delta$ is the reorganization energy. Exact distribution, $\langle\sigma_z\rangle^{eq} = -\tanh(\beta \epsilon)$, (black line), MJ-GQME (green open squares), PBME  (red filled circles, dotted line), Ehrenfest (crosses, navy dashed line), and FSSH (turquoise diamonds, dot-dashed line).}
\label{fig:7}
\end{figure}

An equally important consideration in electron transfer processes is the population distribution that the system relaxes to, which is shown in Fig. \ref{fig:7}. Even at large bias FSSH predicts equal populations in both states in the long time limit. The poor performance of FSSH partially reflects our choice to monitor the populations by transforming the adiabatic coefficients. Previous work has suggested that the long time populations are better captured by basis transforming the adiabatic surface populations \cite{schmidt08}. However, using the surface populations results in incorrect short time diabatic population dynamics even at zero time and gives a poor description of electron transfer rates and population relaxation \cite{subotnik11b,shi13}. Ehrenfest captures the correct trend in the long time populations but underestimates the effect of changing the driving force on the long time population difference. In contrast, despite its earlier failure to describe the long time populations in other nonadiabatic biased regimes with strong system-bath coupling (Figs. \ref{fig:3}, \ref{fig:4} and \ref{fig:5}) PBME gives excellent agreement with the exact long time populations. The success of PBME in this Marcus regime is likely due to the high temperature bath that is known to improve the accuracy of the PBME since the approximated back reaction term becomes less important \cite{kelly11}. MJ-GQME gives exact agreement with the long time populations.

The failure of Ehrenfest and FSSH to obtain accurate long time populations is of considerable concern given that most systems in which one would like to investigate the the mechanism of charge or energy transfer involve multiple states. Since all of the methods employed here preserve the trace of the subsystem RDM, the two level system provides a best case scenario since the population must be in one of the two states so only one detailed balance relation exists. However, in many problems the subsystem contains a large number of states and hence the effect of violating detailed balance would be amplified. In these cases one should therefore be very careful in assessing the importance of the pathways by which the relaxation occurs based on the predictions of methods such as Ehrenfest and FSSH. Although PBME performs slightly better than Ehrenfest in the Marcus regime, in the regimes investigated earlier it performed equally poorly in describing relaxation to the correct distribution. As such its application should be approached with care, particularly in biased systems coupled to low temperature baths. Throughout our testing our MJ-GQME approach has proved to be a highly robust approach that is able to accurately interpolate all the way between disparate regimes ranging from those relevant in energy transfer all the way to strongly coupled electron transfer regimes. The fact that it can achieve this while retaining the flexibility to treat the more general forms of the Hamiltonian to which Ehrenfest, FSSH or PBME can be applied suggests it will prove useful in such problems particularly in cases where it is important to go beyond harmonic baths or linear coupling to the environment.

\section{Conclusions}\label{sec:conc}

We have shown that using the MJ solution to the quantum classical Liouville equation combined with the generalized master equation framework circumvents the problems associated with direct application of the MJ approach alone, which leads to rapidly growing statistical errors with time. The resulting MJ-GQME surface hopping approach has been shown to be highly accurate, and yet can be applied to generate long time dynamics in regimes which are strongly nonadiabatic, biased and exhibit strong system-bath coupling. In these regimes we showed that the competing approaches give poor agreement with exact results. The resulting algorithm was shown to be able to reproduce the Marcus turnover curve and yield accurate long time populations despite lying in a formidable parameter regime for trajectory-based approaches. The MJ-GQME approach can therefore be used to accurately tune between perturbative and non perturbative parameter regimes ranging from electron transfer to electronic energy transfer. In addition, because the evaluation of the kernels is based on semiclassical trajectories, it is also easily adaptable to any form of the system, bath or coupling between them i.e. it is in no way limited to the linear coupling or harmonic bath invoked in the spin boson model studied here.

In addition to accuracy, a major consideration is the relative computational cost. For the highly non adiabatic regimes with strong system-bath coupling shown $\sim 10^{6}$ trajectories were typically required for convergence of the MJ-GQME population dynamics. This is considerably higher than the $\sim 10^{4}$ needed in the FSSH, Ehrenfest and PBME schemes. However, the trajectories that are needed to generate the long time dynamics from the master equation are at least an order magnitude shorter than than those required to directly generate the population dynamics using the MJ-GQME method, due to the fast decay of the memory kernels. Hence, in these regimes where most trajectory based approaches fail to obtain the correct population decay behavior and final populations our MJ-GQME approach is of comparable cost and typically no more than one order of magnitude more expensive. However, in adiabatic regimes with weak system-bath coupling the MJ-GQME kernels can be converged with $\sim 10^{4}$ trajectories and hence, combined with the shorter trajectories needed to converge the kernels, is cheaper than the existing methods. Hence, one only pays an additional computational cost in using MJ-GQME over the other approaches when those approaches give quantitatively and often qualitatively incorrect results, and hence their employment would be questionable.

In summary, combining the rigorous MJ-surface hopping algorithm with the GQME in order to evolve the subsystem RDM offers an accurate alternative to FSSH and mean field theories (such as Ehrenfest, PBME, and LSC-IVR) at a modest extra computational cost. The MJ-GQME approach should therefore provide a powerful tool in the investigation of charge and energy transfer in complex systems.

\section{Acknowledgements}
The authors would like to thank Donal Mac Kernan, Qiang Shi and Tim Berkelbach for helpful discussions. T.E.M acknowledges funding from a Terman fellowship and Stanford start-up funds. A.K. acknowledges a post-doctoral fellowship from the Stanford Center for Molecular Analysis and Design (CMAD).


\begin{thebibliography}{58}
\expandafter\ifx\csname natexlab\endcsname\relax\def\natexlab#1{#1}\fi
\expandafter\ifx\csname bibnamefont\endcsname\relax
  \def\bibnamefont#1{#1}\fi
\expandafter\ifx\csname bibfnamefont\endcsname\relax
  \def\bibfnamefont#1{#1}\fi
\expandafter\ifx\csname citenamefont\endcsname\relax
  \def\citenamefont#1{#1}\fi
\expandafter\ifx\csname url\endcsname\relax
  \def\url#1{\texttt{#1}}\fi
\expandafter\ifx\csname urlprefix\endcsname\relax\def\urlprefix{URL }\fi
\providecommand{\bibinfo}[2]{#2}
\providecommand{\eprint}[2][]{\url{#2}}


\bibitem[{\citenamefont{Marcus}(1964)}]{Marcus64}
\bibinfo{author}{\bibfnamefont{R.A.} \bibnamefont{Marcus}}
  \bibinfo{journal}{Annu. Rev. Phys. Chem.} \textbf{\bibinfo{volume}{15}},
  \bibinfo{pages}{155} (\bibinfo{year}{1964}).

\bibitem[{\citenamefont{Marcus/Sutin}(1985)}]{Marcus85}
\bibinfo{author}{\bibfnamefont{R.A.} \bibnamefont{Marcus}}, and
\bibinfo{author}{\bibfnamefont{N} \bibnamefont{Sutin}},
  \bibinfo{journal}{Biochim. Biophys. Acta} \textbf{\bibinfo{volume}{811}},
  \bibinfo{pages}{265} (\bibinfo{year}{1985}).
  
  \bibitem[{\citenamefont{Bell}(1973)}]{bell73}
\bibinfo{author}{\bibfnamefont{R.~P.} \bibnamefont{Bell}},
  \emph{\bibinfo{title}{The Proton in Chemistry}} (\bibinfo{publisher}{Chapmann
  \& Hall}, \bibinfo{address}{London}, \bibinfo{year}{1973}).

\bibitem[{\citenamefont{Hanna}(2005)}]{hanna05}
\bibinfo{author}{\bibfnamefont{G.} \bibnamefont{Hanna}}
\bibinfo{author}{\bibfnamefont{R.} \bibnamefont{Kapral}}
  \bibinfo{journal}{J. Chem. Phys.} \textbf{\bibinfo{volume}{122}},
  \bibinfo{pages}{244505} (\bibinfo{year}{2005}).

\bibitem[{\citenamefont{Cukier and Nocera}(1998)}]{cukier98}
\bibinfo{author}{\bibfnamefont{R.} \bibnamefont{Cukier}}, \bibnamefont{and}
\bibinfo{author}{\bibfnamefont{D.} \bibnamefont{Nocera}},
  \bibinfo{journal}{Annu. Rev. Phys. Chem.} \textbf{\bibinfo{volume}{49}},
  \bibinfo{pages}{337} (\bibinfo{year}{1998}).

\bibitem[{\citenamefont{Mayer}(2004)}]{mayer04}
\bibinfo{author}{\bibfnamefont{J.~M.} \bibnamefont{Mayer}}
  \bibinfo{journal}{Annu. Rev. Phys. Chem.} \textbf{\bibinfo{volume}{55}},
  \bibinfo{pages}{363} (\bibinfo{year}{2004}).

\bibitem[{\citenamefont{SHS-AS}(2008)}]{shs08}
\bibinfo{author}{\bibfnamefont{S.} \bibnamefont{Hammes-Schiffer}}, \bibnamefont{and}
\bibinfo{author}{\bibfnamefont{A.V.} \bibnamefont{Soudackov}},
  \bibinfo{journal}{J. Phys. Chem. B} \textbf{\bibinfo{volume}{112}},
  \bibinfo{pages}{14108} (\bibinfo{year}{2008}).

\bibitem[{\citenamefont{Engel, Fleming, et al }(2007)}]{engel07}
\bibinfo{author}{\bibnamefont{Engel, G.~S.}};
\bibinfo{author}{\bibnamefont{Calhoun, T.~R.}};
\bibinfo{author}{\bibnamefont{Read, E.~L.}};
\bibinfo{author}{\bibnamefont{Ahn, T.-K.}};
\bibinfo{author}{\bibnamefont{Mancal, T.}};
\bibinfo{author}{\bibnamefont{Cheng, Y.-C.}};
\bibinfo{author}{\bibnamefont{Blankenship, R.~E.}};
\bibinfo{author}{\bibnamefont{Fleming, G.~R.}}
  \textit{\bibinfo{journal}{Nature}} 
  \textbf{\bibinfo{year}{2007}},
  \textit{\bibinfo{volume}{446}},
  \bibinfo{pages}{782-786}.

\bibitem[{\citenamefont{Lee, Cheng and Fleming}(2007)}]{lee07}
\bibinfo{author}{\bibnamefont{Lee, H.}};
\bibinfo{author}{\bibnamefont{Cheng, Y.-C.}};
\bibinfo{author}{\bibnamefont{Fleming, G.~R.}}
  \textit{\bibinfo{journal}{Science}} 
  \textbf{\bibinfo{year}{2007}},
  \textit{\bibinfo{volume}{316}},
  \bibinfo{pages}{1462-1465}.

\bibitem[{\citenamefont{Collini et al}(2009)}]{collini09}
\bibinfo{author}{\bibnamefont{Collini, E.}};
\bibinfo{author}{\bibnamefont{Scholes, G.~D.}}
  \textit{\bibinfo{journal}{Science}}
  \textbf{\bibinfo{year}{2009}},
  \textit{\bibinfo{volume}{323}},
  \bibinfo{pages}{369-373}.

\bibitem[{\citenamefont{Ishizaki and Fleming}(2009)}]{ishizaki09}
\bibinfo{author}{\bibfnamefont{A.} \bibnamefont{Ishizaki}},  \bibnamefont{and}
\bibinfo{author}{\bibfnamefont{G.~R.} \bibnamefont{Fleming}},
 \bibinfo{journal}{Proc. Nat. Acad. Sci.} \textbf{\bibinfo{volume}{106}},
 \bibinfo{pages}{17255} (\bibinfo{year}{2009}).
  
\bibitem[{\citenamefont{Tanimura and Kubo}(1989)}]{tanimura89}
\bibinfo{author}{\bibfnamefont{Y.} \bibnamefont{Tanimura}},  \bibnamefont{and}
\bibinfo{author}{\bibfnamefont{R.} \bibnamefont{Kubo}},
\bibinfo{journal}{J. Phys. Soc. Jpn.} \textbf{\bibinfo{volume}{58}},
\bibinfo{pages}{101} (\bibinfo{year}{1989}).
    
\bibitem[{\citenamefont{Tanimura and Wolynes}(1989)}]{tanimura91}
\bibinfo{author}{\bibfnamefont{Y.} \bibnamefont{Tanimura}},  \bibnamefont{and}
\bibinfo{author}{\bibfnamefont{P.~G.} \bibnamefont{Wolynes}},
\bibinfo{journal}{Phys. Rev. A.} \textbf{\bibinfo{volume}{43}},
\bibinfo{pages}{4131} (\bibinfo{year}{1991}).
  
\bibitem[{\citenamefont{Makri}(1992)}]{makri92}
\bibinfo{author}{\bibfnamefont{N.} \bibnamefont{Makri}}, 
\bibinfo{journal}{Chem. Phys. Lett.} \textbf{\bibinfo{volume}{193}},
\bibinfo{pages}{435} (\bibinfo{year}{1992}).

\bibitem[{\citenamefont{Makrarov and Makri}(1994)}]{makarov94}
\bibinfo{author}{\bibfnamefont{D.~E.} \bibnamefont{Makarov}}, \bibnamefont{and}
\bibinfo{author}{\bibfnamefont{N.} \bibnamefont{Makri}},
\bibinfo{journal}{Chem. Phys. Lett.} \textbf{\bibinfo{volume}{221}},
\bibinfo{pages}{482} (\bibinfo{year}{1994}).

\bibitem[{\citenamefont{Makri and Makrarov}(1994)}]{makri94a}
\bibinfo{author}{\bibfnamefont{N.} \bibnamefont{Makri}}, \bibnamefont{and}
\bibinfo{author}{\bibfnamefont{D.~E.} \bibnamefont{Makarov}},
\bibinfo{journal}{J. Chem. Phys.} \textbf{\bibinfo{volume}{102}},
\bibinfo{pages}{4600} (\bibinfo{year}{1994}).

\bibitem[{\citenamefont{Makri and Makrarov}(1994)}]{makri94b}
\bibinfo{author}{\bibfnamefont{N.} \bibnamefont{Makri}}, \bibnamefont{and}
\bibinfo{author}{\bibfnamefont{D.~E.} \bibnamefont{Makarov}},
\bibinfo{journal}{J. Chem. Phys.} \textbf{\bibinfo{volume}{102}},
 \bibinfo{pages}{4611} (\bibinfo{year}{1994}).
        
\bibitem[{\citenamefont{Egger and Mak}(1994)}]{egger94}
\bibinfo{author}{\bibfnamefont{R.} \bibnamefont{Egger}}, \bibnamefont{and}
\bibinfo{author}{\bibfnamefont{C.} \bibnamefont{Mak}},
\bibinfo{journal}{Phys. Rev. B} \textbf{\bibinfo{volume}{50}},
\bibinfo{pages}{15210} (\bibinfo{year}{1994}).
    
\bibitem[{\citenamefont{Golosov et al.}(1999)}]{golosov99}
\bibinfo{author}{\bibfnamefont{A.~A.} \bibnamefont{Golosov}},
\bibinfo{author}{\bibfnamefont{R.} \bibnamefont{Friesner}},  and
\bibinfo{author}{\bibfnamefont{P.} \bibnamefont{Pechukas}},
  \bibinfo{journal}{J. Chem. Phys.} \textbf{\bibinfo{volume}{110}},
    \bibinfo{pages}{138} (\bibinfo{year}{1999}).

\bibitem[{\citenamefont{Meier and Tannor}(1999)}]{meier99}
\bibinfo{author}{\bibfnamefont{C.}~\bibnamefont{Meier}} \bibnamefont{and}
  \bibinfo{author}{\bibfnamefont{D.~J.}~\bibnamefont{Tannor}}, \bibinfo{journal}{J.
  Chem. Phys.} \textbf{\bibinfo{volume}{111}}, \bibinfo{pages}{3365}
  (\bibinfo{year}{1999}).

\bibitem[{\citenamefont{Xu and Yan}(2002)}]{xu02}
\bibinfo{author}{\bibfnamefont{R.}~\bibnamefont{Xu}} \bibnamefont{and}
  \bibinfo{author}{\bibfnamefont{Y.}~\bibnamefont{Yan}}, \bibinfo{journal}{J.
  Chem. Phys.} \textbf{\bibinfo{volume}{116}}, \bibinfo{pages}{9196}
  (\bibinfo{year}{2002}).

 \bibitem[{\citenamefont{Shi and Geva}(2003)}]{shi03a}
\bibinfo{author}{\bibfnamefont{Q.}~\bibnamefont{Shi}} \bibnamefont{and}
  \bibinfo{author}{\bibfnamefont{E.}~\bibnamefont{Geva}}, \bibinfo{journal}{J.
  Chem. Phys.} \textbf{\bibinfo{volume}{119}}, \bibinfo{pages}{12063}
  (\bibinfo{year}{2003}).

\bibitem[{\citenamefont{Kleinekathofer}(2004)}]{kleinekathofer04}
\bibinfo{author}{\bibfnamefont{U.}~\bibnamefont{Kleinekath\"ofer}}, 
\bibinfo{journal}{J. Chem. Phys.} \textbf{\bibinfo{volume}{121}}, \bibinfo{pages}{2505}
  (\bibinfo{year}{2004}).

\bibitem[{\citenamefont{Muhlbacher and Rabani}(2003)}]{muhlbacher08}
\bibinfo{author}{\bibfnamefont{L.} \bibnamefont{M\"uhlbacher}}, \bibnamefont{and}
\bibinfo{author}{\bibfnamefont{E.} \bibnamefont{Rabani}},
  \bibinfo{journal}{Phys. Rev. Lett.} \textbf{\bibinfo{volume}{100}},
    \bibinfo{pages}{176403} (\bibinfo{year}{2008}).
    
\bibitem[{\citenamefont{Cohen and Rabani}(2011)}]{cohen11}
\bibinfo{author}{\bibfnamefont{G.}~\bibnamefont{Cohen}} \bibnamefont{and}
  \bibinfo{author}{\bibfnamefont{E.}~\bibnamefont{Rabani}}, \bibinfo{journal}{Phys. Rev. B} 
  \textbf{\bibinfo{volume}{84}}, \bibinfo{pages}{075150}
  (\bibinfo{year}{2011}).
  
    \bibitem[{\citenamefont{Kapral}(2006)}]{kapral06}
\bibinfo{author}{\bibfnamefont{R.} \bibnamefont{Kapral}},
  \bibinfo{journal}{Annu. Rev. Phys. Chem.} \textbf{\bibinfo{volume}{57}},
  \bibinfo{pages}{1239} (\bibinfo{year}{2006}).

\bibitem[{\citenamefont{Tully}(2012)}]{Tully12}
\bibinfo{author}{\bibfnamefont{J.} \bibnamefont{Tully}},
  \bibinfo{journal}{J. Chem. Phys.} \textbf{\bibinfo{volume}{137}},
  \bibinfo{pages}{22A301} (\bibinfo{year}{2012}).
 
\bibitem[{\citenamefont{Berkelbach et. al}(2012)}]{berkelbach12}
\bibinfo{author}{\bibfnamefont{T.~C.} \bibnamefont{Berkelbach}},  
\bibinfo{author}{\bibfnamefont{D.~R.} \bibnamefont{Reichman}},\bibnamefont{and}
\bibinfo{author}{\bibfnamefont{T.~E.} \bibnamefont{Markland}},  
  \bibinfo{journal}{J. Chem. Phys.} \textbf{\bibinfo{volume}{136}},
    \bibinfo{pages}{034113} (\bibinfo{year}{2012}).
 
\bibitem[{\citenamefont{Shi and Geva}(2003)}]{shi03c}
\bibinfo{author}{\bibfnamefont{Q.}~\bibnamefont{Shi}} \bibnamefont{and}
  \bibinfo{author}{\bibfnamefont{E.}~\bibnamefont{Geva}}, \bibinfo{journal}{J.
  Chem. Phys.} \textbf{\bibinfo{volume}{118}}, \bibinfo{pages}{8173}
  (\bibinfo{year}{2003}).

\bibitem[{\citenamefont{Poulsen et al}(2003)}]{poulsen03}
\bibinfo{author}{\bibfnamefont{J.~A.}~\bibnamefont{Poulsen}},
\bibinfo{author}{\bibfnamefont{G.}~\bibnamefont{Nyman}}, \bibnamefont{and}
  \bibinfo{author}{\bibfnamefont{P.~J.} \bibnamefont{Rossky}},
  \bibinfo{journal}{J. Chem. Phys.} \textbf{\bibinfo{volume}{119}},
  \bibinfo{pages}{12179} (\bibinfo{year}{2003}).

\bibitem[{\citenamefont{Mclachlan}(1964)}]{mclachlan64}
\bibinfo{author}{\bibfnamefont{A.~D.}~\bibnamefont{McLachlan}}
  \bibinfo{journal}{Mol. Phys.} \textbf{\bibinfo{volume}{8}},
  \bibinfo{pages}{39} (\bibinfo{year}{1964}).
    
  \bibitem[{\citenamefont{Bonella et~al.}(2010)\citenamefont{Bonella, Ciccotti,
  and Kapral.}}]{bonella10}
\bibinfo{author}{\bibfnamefont{S.}~\bibnamefont{Bonella}},
  \bibinfo{author}{\bibfnamefont{G.}~\bibnamefont{Ciccotti}}, \bibnamefont{and}
  \bibinfo{author}{\bibfnamefont{R.}~\bibnamefont{Kapral.}},
  \bibinfo{journal}{Chem. Phys. Lett.} \textbf{\bibinfo{volume}{484}},
  \bibinfo{pages}{399} (\bibinfo{year}{2010}).
 
   \bibitem[{\citenamefont{Sun et~al.}(1998)\citenamefont{Sun, Wang, and
  Miller}}]{sun98}
\bibinfo{author}{\bibfnamefont{X.}~\bibnamefont{Sun}},
  \bibinfo{author}{\bibfnamefont{H.~B.} \bibnamefont{Wang}}, \bibnamefont{and}
  \bibinfo{author}{\bibfnamefont{W.~H.} \bibnamefont{Miller}},
  \bibinfo{journal}{J. Chem. Phys.} \textbf{\bibinfo{volume}{109}},
  \bibinfo{pages}{7064} (\bibinfo{year}{1998}).
 
\bibitem[{\citenamefont{Bonella and Coker}(2005)}]{bonella05}
\bibinfo{author}{\bibfnamefont{S.}~\bibnamefont{Bonella}} \bibnamefont{and}
  \bibinfo{author}{\bibfnamefont{D.~F.} \bibnamefont{Coker}},
  \bibinfo{journal}{J. Chem. Phys.} \textbf{\bibinfo{volume}{122}},
  \bibinfo{pages}{194102} (\bibinfo{year}{2005}).
  
\bibitem[{\citenamefont{Kim et~al.}(2008)\citenamefont{Kim, Nassimi, and
  Kapral.}}]{kim-map08}
\bibinfo{author}{\bibfnamefont{H.}~\bibnamefont{Kim}},
  \bibinfo{author}{\bibfnamefont{A.}~\bibnamefont{Nassimi}}, \bibnamefont{and}
  \bibinfo{author}{\bibfnamefont{R.}~\bibnamefont{Kapral.}},
  \bibinfo{journal}{J. Chem. Phys.} \textbf{\bibinfo{volume}{129}},
  \bibinfo{pages}{084102} (\bibinfo{year}{2008}).

\bibitem[{\citenamefont{Miller}(2012)}]{Miller12}
\bibinfo{author}{\bibfnamefont{W.H.} \bibnamefont{Miller}},
\bibinfo{journal}{J. Chem. Phys.} \textbf{\bibinfo{volume}{136}},
\bibinfo{pages}{210901} (\bibinfo{year}{2012}).
   
 \bibitem[{\citenamefont{Miller}(1970)}]{miller70}
\bibinfo{author}{\bibfnamefont{W.~H.}~\bibnamefont{Miller}},
  \bibinfo{journal}{J. Chem. Phys.} \textbf{\bibinfo{volume}{53}},
  \bibinfo{pages}{3578} (\bibinfo{year}{1970}).
 
 \bibitem[{\citenamefont{Dunkel et~al.}(2008)\citenamefont{Dunkel, Bonella, and
  Coker}}]{dunkel08}
\bibinfo{author}{\bibfnamefont{E.}~\bibnamefont{Dunkel}},
  \bibinfo{author}{\bibfnamefont{S.}~\bibnamefont{Bonella}}, \bibnamefont{and}
  \bibinfo{author}{\bibfnamefont{D.~F.} \bibnamefont{Coker}},
  \bibinfo{journal}{J. Chem. Phys.} \textbf{\bibinfo{volume}{129}},
  \bibinfo{pages}{114106} (\bibinfo{year}{2008}).

\bibitem[{\citenamefont{Huo and Coker}(2011)}]{huo12}
\bibinfo{author}{\bibfnamefont{P.}~\bibnamefont{Huo}} \bibnamefont{and}
  \bibinfo{author}{\bibfnamefont{D.~F.} \bibnamefont{Coker}},
  \bibinfo{journal}{J. Chem. Phys.} \textbf{\bibinfo{volume}{137}},
  \bibinfo{pages}{22A535} (\bibinfo{year}{2012}).

\bibitem[{\citenamefont{Kapral and Ciccotti}(1999)}]{kapral99}
\bibinfo{author}{\bibfnamefont{R.}~\bibnamefont{Kapral}} \bibnamefont{and}
  \bibinfo{author}{\bibfnamefont{G.}~\bibnamefont{Ciccotti}},
  \bibinfo{journal}{J. Chem. Phys.} \textbf{\bibinfo{volume}{110}},
  \bibinfo{pages}{8919} (\bibinfo{year}{1999}).
  
   \bibitem[{\citenamefont{Tully and Preston}(1971)}]{tully71}
\bibinfo{author}{\bibfnamefont{J.~C.} \bibnamefont{Tully}},\bibnamefont{and}
\bibinfo{author}{\bibfnamefont{R.~K.} \bibnamefont{Preston}},
  \bibinfo{journal}{J. Chem. Phys.} \textbf{\bibinfo{volume}{93}},
  \bibinfo{pages}{1061} (\bibinfo{year}{1990}).

 \bibitem[{\citenamefont{Tully}(1990)}]{tully90}
\bibinfo{author}{\bibfnamefont{J.~C.} \bibnamefont{Tully}},
  \bibinfo{journal}{J. Chem. Phys.} \textbf{\bibinfo{volume}{55}},
  \bibinfo{pages}{562} (\bibinfo{year}{1971}).

\bibitem[{\citenamefont{Bittner and Rossky}(1995)}]{bittner95}
\bibinfo{author}{\bibfnamefont{E.~R.} \bibnamefont{Bittner}},\bibnamefont{and}
\bibinfo{author}{\bibfnamefont{P.~J.} \bibnamefont{Rossky}},
  \bibinfo{journal}{J. Chem. Phys.} \textbf{\bibinfo{volume}{103}},
  \bibinfo{pages}{8130} (\bibinfo{year}{1995}).

\bibitem[{\citenamefont{Prezhdo and Rossky}(1997)}]{prezhdo97}
\bibinfo{author}{\bibfnamefont{O.~V.} \bibnamefont{Prezhdo}},\bibnamefont{and}
\bibinfo{author}{\bibfnamefont{P.~J.} \bibnamefont{Rossky}},
  \bibinfo{journal}{J. Chem. Phys.} \textbf{\bibinfo{volume}{107}},
  \bibinfo{pages}{5863} (\bibinfo{year}{1997}).

\bibitem[{\citenamefont{Subotnik and Shenvi}(2011)}]{subotnik11a}
\bibinfo{author}{\bibfnamefont{J.~E.} \bibnamefont{Subotnik}} \bibnamefont{and}
  \bibinfo{author}{\bibfnamefont{N.}~\bibnamefont{Shenvi}},
  \bibinfo{journal}{J. Chem. Phys.} \textbf{\bibinfo{volume}{134}},
  \bibinfo{pages}{024105} (\bibinfo{year}{2011}).
  
   \bibitem[{\citenamefont{Tully}(2012)}]{shuskov12}
\bibinfo{author}{\bibfnamefont{P.} \bibnamefont{Shuskov}},
\bibinfo{author}{\bibfnamefont{R.} \bibnamefont{Li}}, \bibnamefont{and}
\bibinfo{author}{\bibfnamefont{J.~C.} \bibnamefont{Tully}},
  \bibinfo{journal}{J. Chem. Phys.} \textbf{\bibinfo{volume}{137}},
  \bibinfo{pages}{22A549} (\bibinfo{year}{2012}).
   
  \bibitem[{\citenamefont{Landry and Subotnik}(2011)}]{subotnik11b}
  \bibinfo{author}{\bibfnamefont{B.~R.}~\bibnamefont{Landry}}, \bibnamefont{and}
\bibinfo{author}{\bibfnamefont{J.~E.} \bibnamefont{Subotnik}} 
  \bibinfo{journal}{J. Chem. Phys.} \textbf{\bibinfo{volume}{135}},
  \bibinfo{pages}{191101} (\bibinfo{year}{2011}).

\bibitem[{\citenamefont{Landry and Subotnik}(2012)}]{subotnik12}
  \bibinfo{author}{\bibfnamefont{B.~R.}~\bibnamefont{Landry}}, \bibnamefont{and}
\bibinfo{author}{\bibfnamefont{J.~E.} \bibnamefont{Subotnik}} 
  \bibinfo{journal}{J. Chem. Phys.} \textbf{\bibinfo{volume}{137}},
  \bibinfo{pages}{22A513} (\bibinfo{year}{2012}).

\bibitem[{\citenamefont{Bedard-Hearn et al}(2005)}]{bedard-hearn05}
  \bibinfo{author}{\bibfnamefont{M.~J.}~\bibnamefont{Bedard-Hearn}}, 
  \bibinfo{author}{\bibfnamefont{R.~E.}~\bibnamefont{Larsen}}, \bibnamefont{and}
\bibinfo{author}{\bibfnamefont{B.~J.} \bibnamefont{Schwartz}} 
  \bibinfo{journal}{J. Chem. Phys.} \textbf{\bibinfo{volume}{123}},
  \bibinfo{pages}{234106} (\bibinfo{year}{2005}).
 
 \bibitem[{\citenamefont{MacKernan et~al.}(2002)\citenamefont{MacKernan,
  Ciccotti, and Kapral.}}]{mackernan02}
\bibinfo{author}{\bibfnamefont{D.}~\bibnamefont{MacKernan}},
  \bibinfo{author}{\bibfnamefont{G.}~\bibnamefont{Ciccotti}}, \bibnamefont{and}
  \bibinfo{author}{\bibfnamefont{R.}~\bibnamefont{Kapral.}},
  \bibinfo{journal}{J. Phys.: Condens. Matter} \textbf{\bibinfo{volume}{14}},
  \bibinfo{pages}{9069} (\bibinfo{year}{2002}).

 \bibitem[{\citenamefont{MacKernan et~al.}(2008)\citenamefont{MacKernan,
  Ciccotti, and Kapral.}}]{mackernan08}
\bibinfo{author}{\bibfnamefont{D.}~\bibnamefont{MacKernan}},
  \bibinfo{author}{\bibfnamefont{R.}~\bibnamefont{Kapral.}},\bibnamefont{and}
  \bibinfo{author}{\bibfnamefont{G.}~\bibnamefont{Ciccotti}}, 
  \bibinfo{journal}{J. Phys. Chem. B} \textbf{\bibinfo{volume}{112}},
  \bibinfo{pages}{424} (\bibinfo{year}{2008}).

\bibitem[{\citenamefont{Kelly and Kapral}(1999)}]{kelly10}
  \bibinfo{author}{\bibfnamefont{G.}~\bibnamefont{Ciccotti}}, \bibnamefont{and}
\bibinfo{author}{\bibfnamefont{R.}~\bibnamefont{Kapral}} 
  \bibinfo{journal}{J. Chem. Phys.} \textbf{\bibinfo{volume}{133}},
  \bibinfo{pages}{084502} (\bibinfo{year}{2010}).
  
  \bibitem[{\citenamefont{Nakajima}(1958)}]{nakajima58}
\bibinfo{author}{\bibfnamefont{S.} \bibnamefont{Nakajima}},
  \bibinfo{journal}{Prog. Theor. Phys.} \textbf{\bibinfo{volume}{20}},
  \bibinfo{pages}{948} (\bibinfo{year}{1958}).

\bibitem[{\citenamefont{Zwanzig}(1960)}]{zwanzig60}
\bibinfo{author}{\bibfnamefont{R.} \bibnamefont{Zwanzig}},
  \bibinfo{journal}{J. Chem. Phys.} \textbf{\bibinfo{volume}{33}},
  \bibinfo{pages}{1338} (\bibinfo{year}{1960}).
 
\bibitem[{\citenamefont{Shi and Geva}(2004)}]{shi04a}
\bibinfo{author}{\bibfnamefont{Q.}~\bibnamefont{Shi}} \bibnamefont{and}
  \bibinfo{author}{\bibfnamefont{E.}~\bibnamefont{Geva}}, \bibinfo{journal}{J.
  Chem. Phys.} \textbf{\bibinfo{volume}{120}}, \bibinfo{pages}{10647}
  (\bibinfo{year}{2004}).

\bibitem[{\citenamefont{Shi and Geva}(2004)}]{shi04b}
\bibinfo{author}{\bibfnamefont{Q.}~\bibnamefont{Shi}} \bibnamefont{and}
\bibinfo{author}{\bibfnamefont{E.}~\bibnamefont{Geva}}, \bibinfo{journal}{J.
 Chem. Phys.} \textbf{\bibinfo{volume}{121}}, \bibinfo{pages}{3393}
(\bibinfo{year}{2004}).

\bibitem[{\citenamefont{Zhang et al}(2006)}]{geva06}
\bibinfo{author}{\bibfnamefont{M.-L.}~\bibnamefont{Zhang}},
\bibinfo{author}{\bibfnamefont{B. J.}~\bibnamefont{Ka}} \bibnamefont{and}
  \bibinfo{author}{\bibfnamefont{E.}~\bibnamefont{Geva}}, \bibinfo{journal}{J.
  Chem. Phys.} \textbf{\bibinfo{volume}{125}}, \bibinfo{pages}{044106}
  (\bibinfo{year}{2006}).
  
\bibitem[{\citenamefont{Leggett}(1987)}]{leggett87}
\bibinfo{author}{\bibfnamefont{A.} \bibnamefont{Leggett}},
\bibinfo{author}{\bibfnamefont{S.} \bibnamefont{Chakravarty}},
\bibinfo{author}{\bibfnamefont{A.} \bibnamefont{Dorsey}},
\bibinfo{author}{\bibfnamefont{M.} \bibnamefont{Fisher}},
\bibinfo{author}{\bibfnamefont{A.} \bibnamefont{Garg}},  \bibnamefont{and}
\bibinfo{author}{\bibfnamefont{R.} \bibnamefont{Zwerger}}, 
  \bibinfo{journal}{Rev. Mod. Phys.} \textbf{\bibinfo{volume}{59}},
  \bibinfo{pages}{1} (\bibinfo{year}{1987}).

 \bibitem[{\citenamefont{Weiss}(1992)}]{weiss92}
\bibinfo{author}{\bibfnamefont{U.} \bibnamefont{Weiss}},
  \emph{\bibinfo{title}{Quantum Dissipative Systems}} (\bibinfo{publisher}{World Scientific}, \bibinfo{address}{Singapore}, \bibinfo{year}{1992}).

\bibitem[{\citenamefont{Imre et~al.}(1967)\citenamefont{Imre, \"{O}zizmir,
  Rosenbaum, and Zwiefel}}]{imre67}
\bibinfo{author}{\bibfnamefont{K.}~\bibnamefont{Imre}},
  \bibinfo{author}{\bibfnamefont{E.}~\bibnamefont{\"{O}zizmir}},
  \bibinfo{author}{\bibfnamefont{M.}~\bibnamefont{Rosenbaum}},
  \bibnamefont{and} \bibinfo{author}{\bibfnamefont{P.~F.}
  \bibnamefont{Zwiefel}}, \bibinfo{journal}{J. Math. Phys.}
  \textbf{\bibinfo{volume}{5}}, \bibinfo{pages}{1097} (\bibinfo{year}{1967}).
  
\bibitem[{\citenamefont{Tao and Miller}(2010)}]{tao10}
\bibinfo{author}{\bibfnamefont{G.}~\bibnamefont{Tao}} \bibnamefont{and}
\bibinfo{author}{\bibfnamefont{W.~H.} \bibnamefont{Miller}},
\bibinfo{journal}{J. Phys. Chem. Lett.} \textbf{\bibinfo{volume}{1}},
\bibinfo{pages}{891} (\bibinfo{year}{2010}).

\bibitem[{\citenamefont{Kelly and Rhee}(2011)}]{kelly11}
\bibinfo{author}{\bibfnamefont{A.}~\bibnamefont{Kelly}} \bibnamefont{and}
\bibinfo{author}{\bibfnamefont{Y.~M.} \bibnamefont{Rhee}},
\bibinfo{journal}{J. Phys. Chem. Lett.} \textbf{\bibinfo{volume}{2}},
\bibinfo{pages}{808} (\bibinfo{year}{2011}).

\bibitem[{\citenamefont{Kelly Van Zon Schofield and Kapral}(2012)}]{kelly12}
  \bibinfo{author}{\bibfnamefont{A.}~\bibnamefont{Kelly}}, 
    \bibinfo{author}{\bibfnamefont{R.}~\bibnamefont{van Zon}},
      \bibinfo{author}{\bibfnamefont{J.}~\bibnamefont{Schofield}}, \bibnamefont{and}
\bibinfo{author}{\bibfnamefont{R.}~\bibnamefont{Kapral}} 
  \bibinfo{journal}{J. Chem. Phys.} \textbf{\bibinfo{volume}{136}},
  \bibinfo{pages}{084101} (\bibinfo{year}{2012}).
  
  \bibitem[{\citenamefont{Ananth et~al.}(2007)\citenamefont{Ananth, Venkataraman,
 and Miller}}]{ananth07}
\bibinfo{author}{\bibfnamefont{N.}~\bibnamefont{Ananth}},
\bibinfo{author}{\bibfnamefont{C.}~\bibnamefont{Venkataraman}}, \bibnamefont{and}
\bibnamefont{and} \bibinfo{author}{\bibfnamefont{W.~H.}
 \bibnamefont{Miller}}, \bibinfo{journal}{J. Chem. Phys.}
 \textbf{\bibinfo{volume}{127}}, \bibinfo{pages}{084114}
 (\bibinfo{year}{2007}).

  \bibitem[{\citenamefont{Muller Stock}(1997)}]{stock97}
\bibinfo{author}{\bibfnamefont{U.}~\bibnamefont{M\"uller}} \bibnamefont{and}
\bibinfo{author}{\bibfnamefont{G.}~\bibnamefont{Stock}}
\bibinfo{journal}{J. Chem. Phys.} \textbf{\bibinfo{volume}{107}},
\bibinfo{pages}{6230} (\bibinfo{year}{1997}).
  
  \bibitem[{\citenamefont{martinez}(2007)}]{martinez07}
\bibinfo{author}{\bibfnamefont{M.} \bibnamefont{Ben-Nun}},\bibnamefont{and}
\bibinfo{author}{\bibfnamefont{T.~J.} \bibnamefont{Mart\'inez}},
  \bibinfo{journal}{Isr. J. Chem.}, \textbf{\bibinfo{volume}{47}},
  \bibinfo{pages}{75-88} (\bibinfo{year}{2007}).
  
   \bibitem[{\citenamefont{Zusman}(1980)}]{Zus80}
\bibinfo{author}{\bibfnamefont{L.~D.} \bibnamefont{Zusman}},  
  \bibinfo{journal}{Chem. Phys.} \textbf{\bibinfo{volume}{49}},
    \bibinfo{pages}{295} (\bibinfo{year}{1980}).
    
 \bibitem[{\citenamefont{Garg et al}(1985)}]{Garg85}
\bibinfo{author}{\bibfnamefont{A.} \bibnamefont{Garg}},  
\bibinfo{author}{\bibfnamefont{J.~N.} \bibnamefont{Onuchic}}, \bibnamefont{and}
\bibinfo{author}{\bibfnamefont{V.}~\bibnamefont{Ambegaokar}},
  \bibinfo{journal}{J. Chem. Phys.} \textbf{\bibinfo{volume}{83}},
    \bibinfo{pages}{4491} (\bibinfo{year}{1985}).

\bibitem[{\citenamefont{Muhlbacher and Egger}(2004)}]{Muhlbacher04}
\bibinfo{author}{\bibfnamefont{L.} \bibnamefont{M\"uhlbacher}}, \bibnamefont{and}
\bibinfo{author}{\bibfnamefont{R.} \bibnamefont{Egger}},
  \bibinfo{journal}{Chem. Phys.} \textbf{\bibinfo{volume}{296}},
    \bibinfo{pages}{193} (\bibinfo{year}{2004}).

 \bibitem[{\citenamefont{Schmidt}(2008)}]{schmidt08}
\bibinfo{author}{\bibfnamefont{J.~R.} \bibnamefont{Schmidt}},
\bibinfo{author}{\bibfnamefont{P.~V.} \bibnamefont{Parandekar}}, \bibnamefont{and}
\bibinfo{author}{\bibfnamefont{J.~C.} \bibnamefont{Tully}},
  \bibinfo{journal}{J. Chem. Phys.} \textbf{\bibinfo{volume}{129}},
  \bibinfo{pages}{044104} (\bibinfo{year}{2008}).

 \bibitem[{\citenamefont{Shi}(2013)}]{shi13}
\bibinfo{author}{\bibfnamefont{W.} \bibnamefont{Xie}},
\bibinfo{author}{\bibfnamefont{S.} \bibnamefont{Bai}},
\bibinfo{author}{\bibfnamefont{L.} \bibnamefont{Zhu}}, \bibnamefont{and}
\bibinfo{author}{\bibfnamefont{Q.} \bibnamefont{Shi}},
  \bibinfo{journal}{J. Phys. Chem. A},
  \bibinfo{pages}{jp400462f} (\bibinfo{year}{2013}).
  
\end{thebibliography}
\end{document}